\documentclass[aps,preprint,superscriptaddress,groupedaddress,nofootinbib]{revtex4}

\usepackage{graphicx}
\usepackage[normalem]{ulem}
\usepackage[export]{adjustbox}
\usepackage{float}
\usepackage{amsmath}
\usepackage{amssymb}
\usepackage{booktabs} % To thicken table lines
\usepackage{hhline}
\usepackage{multirow}
\usepackage[rightcaption]{sidecap}
\usepackage{hyperref}
\usepackage{xcolor}
\graphicspath{ {./Figures/} }

\def\beq{\begin{equation}}
\def\eeq{\end{equation}}
\def\bea{\begin{eqnarray}}
\def\eea{\end{eqnarray}}

\newcommand{\vw}{v_w}

\def\figureautorefname~#1\null{Fig.\,#1\null}
\def\tableautorefname~#1\null{Tab.\,#1\null}

\def\equationautorefname~#1\null{Eq.\,(#1)\null}
\allowdisplaybreaks[4]

\begin{document}

%%%%%%%%%%%%%%%%%%%%%%%%%%%%%%%%%%%%%%%%%%%%%%%%%%%%%%%
%%%%%%%%%%%%%%%%       Title Page    %%%%%%%%%%%%%%%%%%%%%%%%%%%%%%%
%%%%%%%%%%%%%%%%%%%%%%%%%%%%%%%%%%%%%%%%%%%%%%%%%%%%%%%
\title{Gravitational Waves, Bubble Profile, and Baryon Asymmetry in the Complex 2HDM}

%authors
\author{Dorival Gon\c{c}alves}
\email{dorival@okstate.edu}
\affiliation{Department of Physics, Oklahoma State University, Stillwater, OK, 74078, USA}
\author{Ajay Kaladharan}
\email{kaladharan.ajay@okstate.edu}
\affiliation{Department of Physics, Oklahoma State University, Stillwater, OK, 74078, USA}
\author{Yongcheng Wu}
\email{ycwu@njnu.edu.cn}
\affiliation{Department of Physics and Institute of Theoretical Physics, Nanjing Normal University, Nanjing, 210023, China}
\affiliation{Department of Physics, Oklahoma State University, Stillwater, OK, 74078, USA}

% e-mail addresses: one for each author, in the same order as the authors
% \emailAdd{dorival@okstate.edu}
% \emailAdd{kaladharan.ajay@okstate.edu}
% \emailAdd{ycwu@njnu.edu.cn}

% \abstract{
\begin{abstract}
    This study explores the generation of the observed baryon asymmetry of the Universe within the complex Two Higgs Doublet Model (C2HDM) while considering theoretical and current experimental constraints. In our investigation, we analyze critical elements of the Higgs potential to understand the phase transition pattern.  Specifically, we examine the formation of the barrier and the uplifting of the true vacuum state, which play crucial roles in facilitating a strong first-order phase transition. Furthermore, we explore the potential gravitational wave signals associated with this phase transition pattern and investigate the parameter space points that can be probed with LISA. Finally, we compare the impact of different approaches to describing the bubble profile on the calculation of the baryon asymmetry. We contrast the typically used kink profile approximation against the explicit solution of the tunneling profile. We find that a non-negligible range of the C2HDM parameter space results in significant discrepancies in the baryon asymmetry estimation between these two approaches. Through an examination of the parameter space, we identify a benchmark point that satisfies the observed baryon asymmetry.
\end{abstract}
% }

%\keywords{Beyond Standard Model, Phenomenological Models, Higgs Physics, Top physics, LHC}

%%%%%%%%%%%%%%%%%%%%%%%%%%%%%%%%%%%%%%%%%

\maketitle
\flushbottom
\clearpage

%%%%%%%%%%%%%%%%%%%%%%%%%%%%%%%%%%%%%%%%%%%%%%%%%%%%%%%%%%%
\section{Introduction}
\label{sec:intro}

Understanding the origin of the matter-antimatter asymmetry of the Universe, known as the baryon asymmetry of the Universe (BAU), is a fundamental question in particle physics and cosmology. The  asymmetry between baryons and antibaryons in the early Universe can be quantitatively evidenced through the
baryon-to-entropy ratio measurement $n_B/s\simeq 8.6\times 10^{-11}$~\cite{Planck:2018vyg}, exceeding the expected value for a symmetric scenario by several orders of magnitude. As a consequence, the majority of antibaryons underwent annihilation during the thermal history, leaving behind a significant density of baryons in the present Universe. The essential ingredients required for generating this baryon asymmetry are theoretically well understood and encapsulated by the three Sakharov conditions~\cite{Sakharov}. These conditions demand the violation of baryon number, the presence of $C$ and $CP$ violation, and a departure from thermal equilibrium. While the Standard Model (SM) satisfies the requirements for baryon number violation and $C$ violation, it falls short in providing a sufficiently robust source of $CP$ violation. Additionally, the observed Higgs mass of $m_h=125$~GeV precludes the necessary out-of-equilibrium conditions through a strong first-order phase transition~\cite{Huet:1994jb,Kajantie:1996mn}. Thus, the quest for baryogenesis requires physics beyond the SM~\cite{Trodden:1998ym,Cohen:1993nk,Carena:1996wj,Morrissey:2012db}.

Among possible extensions, the complex Two-Higgs Doublet Model (C2HDM) can potentially provide both of the missing ingredients: strong first-order electroweak phase transition  and additional sources of CP-violation~\cite{Kuzmin:1985mm,Basler:2017uxn,Basler:2019iuu}. In this work, we explore the phase transition pattern and the feasibility of generating the observed baryon asymmetry within the context of the C2HDM. Central to our investigation is the shape of the Higgs potential, which plays a crucial role in determining the nature of the phase transition. We focus on the formation of the barrier and the upliftment of the true vacuum state, as these factors are instrumental in driving the phase transition from a smooth crossover to a strong first-order transition. Our analysis builds upon previous studies for other new physics extensions~\cite{Dorsch:2017nza,EWPT-NMSSM,EWPT-Nature,Goncalves:2021egx}, where it was observed that the intensity of the phase transition is closely linked to the elevation of the true vacuum relative to the symmetric one at zero temperature. The prevalence of one-loop effects over thermal corrections, particularly when $\xi_c>1$, enhances the strength of the phase transition~\cite{Goncalves:2021egx}. However, it should be noted that if the one-loop correction is too large, the universe may become trapped in the electroweak symmetric vacuum, resulting in an incomplete phase transition~\cite{Goncalves:2021egx,Biekotter:2022kgf}. Consequently, as we will show, a significant portion of parameter points with large $\xi_c$ values become unphysical in this scenario.

The first-order phase transition in the early Universe can generate stochastic gravitational waves (GW)  whose characteristic peak frequency is associated with the phase transition temperature. After redshifting to the present time, the GW spectrum would have a peak frequency at the mHz range for the phase transition at the electroweak scale~\cite{Grojean:2006bp,Athron:2023xlk}. This presents an exciting prospect to probe electroweak phase transition (EWPT) at LISA~\cite{Caprini:2019egz}, designed to be sensitive to mHz frequency signals. Hence, we also investigate the parameter space points in C2HDM that can be probed using LISA.

Through an extensive exploration of the parameter space, we note that the C2HDM can describe the observed baryon asymmetry, although only for a limited set of parameter space points. In this regard, we compare two different approaches to describe the bubble profile, a key ingredient in the BAU estimation. The commonly adopted kink profile parameterization~\cite{Bodeker:2004ws,Fromme:2006wx, Fromme:2006cm,Basler:2020nrq, Basler:2021kgq} and the explicit solution for the tunneling equation are examined to assess their impact on the resulting baryon asymmetry. Our analysis reveals relevant deviations between the BAU calculation between these two approaches. While the majority of parameter space points yield similar results using both methods, a notable fraction exhibits significant differences, sometimes varying by several orders of magnitude. To understand these discrepancies, we scrutinize the behavior of the source term in front of the bubble wall, which sheds light on the distinct asymmetry values obtained from the two profile assumptions.

The paper is organized as follows. In \autoref{sec:model}, we provide a brief overview of the complex Two Higgs Double Model. \autoref{sec:Veff} discusses the one-loop finite temperature effective potential. It is followed by a discussion on electroweak phase transition and GW signals in~\autoref{sec:EWPT_GW}. In \autoref{sec:upliftment}, we study how the shape of the Higgs potential will affect the EWPT, focusing on the barrier formation and the vacuum upliftment. In \autoref{sec:BAU}, we present the details for the baryon asymmetry calculation. The results of the BAU are presented in~\autoref{sec:BassymC2HDM}, where we also contrast the results depending on the bubble profile estimation. Finally, we summarize in \autoref{sec:summary}. Details of the  parameterization for the C2HDM scan are presented in~\autoref{app:parameters}.

%%%%%%%%%%%%%%%%%%%%%%%%%%%%%%%%%%%%%%%%%%%%%%%%%%%%%%%%%%%
\section{Complex Two Higgs Doublet Model}
\label{sec:model}
The two Higgs doublet  model (2HDM) lays out a compelling extension of the SM in line with current experimental constraints ~\cite{Branco:2011iw}. This work considers CP-violating 2HDM with a softly broken $\mathbb{Z}_2$ symmetry. Within this framework, the tree-level potential is given by
%----
\begin{align}
    V_{0}(\Phi_1,\Phi_2) =& m_{11}^2\Phi_1^\dagger\Phi_1 + m_{22}^2\Phi_2^\dagger\Phi_2 - (m_{12}^2\Phi_1^\dagger\Phi_2 + h.c.) + \frac{\lambda_1}{2}(\Phi_1^\dagger\Phi_1)^2 + \frac{\lambda_2}{2}(\Phi_2^\dagger\Phi_2)^2\nonumber \\
            & + \lambda_3(\Phi_1^\dagger\Phi_1)(\Phi_2^\dagger\Phi_2) + \lambda_4(\Phi_1^\dagger\Phi_2)(\Phi_2^\dagger\Phi_1) + \left(\frac{\lambda_5}{2}(\Phi_1^\dagger\Phi_2)^2 + h.c.\right),
\label{equ:v_tree}
\end{align}
%----
where the mass term $m_{12}^2$ and quartic coupling $\lambda_5$ are complex and all other mass terms and quartic couplings are taken to be real. However, one of the phases in $m_{12}^2$ and $\lambda_5$ can be removed by a phase redefinition of $\Phi_2$. In this work, we always keep $m_{12}^2$ real and $\lambda_5$ will be complex at zero temperature. Hence, overall in such setup, there is only one independent physical CP violation phase. To preclude dangerous tree-level Flavor Changing Neutral Currents (FCNC)~\cite{PhysRevD.15.1958,PhysRevD.15.1966}, we impose a  $\mathbb{Z}_2$ symmetry softly broken by the  $m_{12}^2$ term, under which $\Phi_1\to \Phi_1$ and $\Phi_2\to -\Phi_2$. Following electroweak symmetry breaking, the neutral components of $\Phi_1$ and $\Phi_2$ develop non-zero vacuum expectation values (VEVs).

Expanding around the  VEVs ${\omega}_i$, the scalar doublets $\Phi_i$ can be written as
{
\begin{align}
\Phi_1 = \begin{pmatrix}
H_1^+\\
\frac{\omega_1 + H_1^0 + i A_1^0}{\sqrt{2}}
\end{pmatrix}
\hspace{0.5cm} \text{and} \hspace{0.5cm}
\Phi_2 = e^{i\omega_\theta}\begin{pmatrix}
    H_2^+ + \frac {\omega_{\rm CB}}{\sqrt{2}}\\
    \frac{\omega_2 + H_2^0 + i A_2^0}{\sqrt{2}}
\end{pmatrix}
\label{eq:doublet}
\end{align}
}
%----
where at zero temperature VEVs $v_i\equiv {\omega}_i|_{T=0}$, $i=1,2$ are linked to SM VEV by $v_1^2+v_2^2=v^2\approx (246~{\rm GeV})^2$.
Whereas an additional source of CP-violation should decrease at zero temperature $({\omega}_{\theta}|_{T=0}\to 0)$ to comply with the stringent electric dipole moment (EDM) constraints~\cite{ACME:2018yjb}, the dynamical generation of CP-violation at high temperatures offers a potential avenue for a CP-violating mechanism crucial to the success of Electroweak baryogenesis.
To account for a more comprehensive scenario, we also incorporate a possible charge-breaking at high temperature, $\omega_{\mathrm{CB}}$. Since a non-zero charge-breaking VEV at zero temperature would lead to massive photons, we impose $v_{CB}=0$.

The  scalar sector in the CP-violating 2HDM has five  physical mass eigenstates: three CP-mixed neutral scalars $H_i$ and one charged scalar pair $H^\pm$. The correspondence between mass eigenstates and gauge eigenstates is established by the mixing angle $\beta$ in CP-odd and charged sectors and another three angles {$\alpha, \; \alpha_b$ and $\alpha_c$} mixing the CP-odd and CP-even scalars:
%-------------------------------------------------------
{\begin{align}
    \left(\begin{array}{c}
        G^\pm\\
        H^\pm
    \end{array}\right) =
    \left(\begin{array}{cc}
        c_\beta  & s_\beta \\
        -s_\beta & c_\beta
    \end{array}\right)\,
    \left(\begin{array}{c}
        H_1^\pm\\
        H_2^\pm
    \end{array}\right),\,\,\,
    \left(\begin{array}{c}
        G^0\\
        A
    \end{array}\right)= \left(\begin{array}{cc}
        c_\beta  & s_\beta \\
        -s_\beta & c_\beta
    \end{array}\right)\
    \left(\begin{array}{c}
        A_1^0\\
        A_2^0
    \end{array}\right),
\end{align}}
%------------------------------------------------------
{\begin{align}
    \left(\begin{array}{c}
        H_1\\
        H_2\\
        H_3
    \end{array}\right) = O\begin{pmatrix}
        H_1^0\\
        H_2^0\\
        A
    \end{pmatrix}=
    \begin{pmatrix}
        -s_\alpha c_{\alpha_b} & c_\alpha c_{\alpha_b} & s_{\alpha_b}\\
        c_\alpha c_{\alpha_c} + s_\alpha s_{\alpha_b}s_{\alpha_c} & s_\alpha c_{\alpha_c} - c_\alpha s_{\alpha_b} s_{\alpha_c} & c_{\alpha_b}s_{\alpha_c} \\
        -c_\alpha s_{\alpha_c} + s_\alpha s_{\alpha_b}c_{\alpha_c} & -s_\alpha s_{\alpha_c} - c_\alpha s_{\alpha_b} c_{\alpha_c} & c_{\alpha_b} c_{\alpha_c}
\end{pmatrix}
    \left(\begin{array}{c}
        H_1^0\\
        H_2^0\\
        A
    \end{array}\right).
    \label{eq:betamixing}
\end{align}
The mixing angle $\beta$ is defined as $t_\beta \equiv \tan\beta = v_2/v_1$ ($\cos\beta \equiv c_\beta$ $\sin\beta \equiv s_\beta$). We also define $c_x \equiv \cos x$ and $s_x \equiv \sin x$.}

{At zero temperature, the physical parameters in the scalar sector include the VEVs ($v_1=vc_\beta,v_2=vs_\beta,\theta\equiv\langle\omega_\theta\rangle$), the masses of the scalar eigenstates ($m_{H_i}$ and $m_{H^\pm}$), the mixing angles ($\alpha$, $\alpha_b$, and $\alpha_c$), and $m_{12}^2$. Note that, as we mentioned earlier, there is only one physical CP violation phase, \emph{i.e.}, only one of $\theta$, $\alpha_b$, and $\alpha_c$ is independent. In this work, we keep $\alpha_c$ as an independent input while calculating $\theta$ and $\alpha_b$ from other parameters. Hence, we choose the input parameters to be
\begin{align}
    \label{equ:inputs}
    v = 246\,{\rm GeV},\, t_\beta,\, c_{\beta-\alpha},\, \alpha_c,\, m_{12}^2,\, m_h = 125\,{\rm GeV},\, m_{H_\uparrow}, m_{H_\downarrow},\,m_{H^\pm}\,,
\end{align}
which match the 9 real parameters in the potential~\autoref{equ:v_tree}. Here, $m_{H_\uparrow}$ and $m_{H_\downarrow}$ represent the masses of heavier and lighter beyond the Standard Model (BSM) neutral scalars, respectively. The detailed mapping between the parameters in~\autoref{equ:inputs} and those in~\autoref{equ:v_tree} can be found in~\autoref{app:parameters}. This parameterization for the CP-violating 2HDM is similar to the scan performed for CP-conserving 2HDM in our earlier works~\cite{Goncalves:2021egx,Goncalves:2022wbp} in the sense that it provides the scans over all physical BSM scalar masses ($m_{H_\uparrow}$, $m_{H_\downarrow}$ and $m_{H^\pm}$) and CP-violating angle ($\alpha_c$).}
The phase transition pattern in 2HDM, to a large extent, depends on the masses of additional scalars and corresponding mass splittings~\cite{Goncalves:2021egx}. Hence, the numerical scan performed over three scalar masses is more suitable than one of the scalar masses written as a function of other scan variables.

Within the Yukawa sector, there are four distinct $\mathbb{Z}_2$ charge assignments that effectively preclude tree-level FCNC. In this study, we focus on two specific scenarios: type-I and type-II. In the type-I scenario, all fermions exclusively couple with $\Phi_2$, while in the type-II scenario, only up quarks couple with $\Phi_2$, with down quarks and charged leptons coupling with $\Phi_1$. To thoroughly explore these possibilities, we conduct a random uniform scan, encompassing both type-I and type-II configurations, over the parameter space region
%----
\begin{align}
    \tan\beta &\in (0.8,25)\,,                &m_{12}^2&\in(10^{-3},5\times 10^5)\,{\rm GeV}^2\,,  &m_{H_{\uparrow/\downarrow}}&\in(30,1500)\rm\,GeV \,,
        % \label{eq:param_scan1}
    \nonumber \\
    \alpha_c&\in(-\frac {\pi}{2},\frac{\pi}{2})\,,& \cos&(\beta-\alpha)\in(-0.3,0.3)\,,                     &m_{H^\pm}&\in(150,1500)\,{\rm GeV}.
     \label{eq:param_scan}
\end{align}
%----

We performe the parameter space scan by implementing the parametrization detailed in~\autoref{app:parameters} in {\tt ScannerS}~\cite{Coimbra:2013qq,Muhlleitner:2020wwk}. Using {\tt ScannerS}, we impose constraints from perturbative unitarity~\cite{Lee:1977eg,Kanemura:1993hm,Ginzburg:2005dt}, boundedness from below~\cite{Ivanov:2018jmz}, vacuum stability~\cite{Hollik:2018wrr,Ferreira:2019iqb}, electroweak precision,  and flavor constraints.
EDM constraints are also imposed using the stringent limits from the ACME collaboration~\cite{ACME:2018yjb}. Furthermore, constraints from the 125~GeV Higgs boson measurements and additional scalar searches are carried out using  {\tt HiggsBounds} and {\tt HiggsSignals}~\cite{Bechtle:2020pkv,Bechtle:2020uwn,Atkinson:2022pcn}.

%%%%%%%%%%%%%%%%%%%%%%%%%%%%%%%%%%%%%%%%%%%%%%%%%%%%%%%%%%%
\section{One-loop Finite Temperature Effective Potential}
\label{sec:Veff}
%%%%%%%%%%%%%%%%%%%%%%%%%%%%%%%%%%%%%%%%%%%%%%%%%%%%%%%%%%%
We use loop-corrected finite temperature effective potential to determine the dynamics of electroweak symmetry breaking in the early Universe. Along with the tree-level potential $V_0$ from~\autoref{equ:v_tree}, we also include the Coleman-Weinberg potential $V_{\rm CW}$ and counterterms $V_{\rm CT}$ that encode one-loop corrections at zero temperature, and finite-temperature corrections $V_T$. The effective potential is given by
%-------
\begin{align}
    V_{\rm eff} = V_0 + V_{\rm CW} + V_{\rm CT}+ V_T\,.
    \label{eq:Veff}
\end{align}
%-------

The Coleman-Weinberg potential in the Landau gauge can be written, using $\overline{\mathrm{MS}}$ renormalization prescription as~\cite{PhysRevD.7.1888}
%-------
\begin{align}
    V_{\rm CW} &= \sum_i \frac{n_i}{64\pi^2}m_i^4(\Phi_1,\Phi_2)\left[\log \left(\frac{m_i^2(\Phi_1,\Phi_2)}{\mu^2}\right)-c_i\right]\,,
    \label{eq:CW}
\end{align}
%-------
where the index $i$ runs over all particles in the thermal bath with field-dependent mass $m_i(\Phi_1,\Phi_2)$, including Higgs bosons, massive gauge bosons, Goldstone bosons, longitudinal photon, and fermions. The parameter $n_i$ represents the number of degrees of freedom for each particle, with $n_i>0$ for bosons and $n_i<0$ for fermions.  In the $\overline{\text{MS}}$ renormalization procedure, the coefficient $c_i$ takes the value of $5/6$ for gauge bosons and $3/2$ otherwise. Moreover, we set the renormalization scale $\mu$ to the zero-temperature VEV, $\mu=v(T=0)\approx 246~$GeV.\footnote{A renormalization group improved calculation can be taken into account for a further refined estimation~\cite{Chiang:2017nmu}. For the renormalization scale $\mu^2$ dependence of effective potential at finite temperature, we refer to Ref.~\cite{Gould:2021oba}.}

The one-loop effects of the Coleman-Weinberg potential result in shifts of the mixing angle and scalar masses from their tree-level values. To perform a consistent parameter scan, we adopt an on-shell renormalization scheme, which enforces the parameters to match their tree-level values~\cite{Camargo-Molina:2016moz, Basler:2016obg, Basler:2021kgq}, by proper counterterms determined according to
%-------
\begin{align}
\partial_{\phi_i}(V_\text{CW}+V_\text{CT})|_{\omega=\omega_\text{tree}}=0 \,,
\label{eq:ren1}\\
\partial_{\phi_i}\partial_{\phi_j}(V_\text{CW}+V_\text{CT})|_{\omega=\omega_\text{tree}}=0\, ,
\label{eq:ren2}
\end{align}
%-------
where $\phi_i$ ($i=1,...,8$) represents scalar components from the $\Phi_1$ and $\Phi_2$ doublets, $\omega$ denotes the $\omega_i$ values, and $\omega_\text{tree}$ characterizes the minimum of the tree-level potential for the fields in $\Phi_1$ and $\Phi_2$. The first and second derivatives of $V_{\rm CW}$ are consistently defined with an analytical expression in Ref.~\cite{Camargo-Molina:2016moz}.\footnote{{The first and second derivatives of the effective potential exhibit IR divergence at zero temperature as a result of the contribution from Goldstone bosons~\cite{Elias-Miro:2014pca, Martin:2014bca}. To address this, we use the analytical expression provided in Ref.~\cite{Camargo-Molina:2016moz} to evaluate derivatives of effective potential and compute the counterterms.}} The first renormalization condition, given by~\autoref{eq:ren1}, ensures that the minimum of the effective potential is not shifted from tree-level minimum, and the second condition, shown in~\autoref{eq:ren2}, guarantees that mixing angles and scalar masses remain the same as their tree-level values.

The one-loop thermal correction $V_T$ in~\autoref{eq:Veff} is given by~\cite{Arnold:1992rz}
%-------
\begin{align}
    V_T&=\frac{T^4}{2\pi^2} \left[
    \sum_f n_f J_+\left(\frac{m_f^2}{T^2}\right)+
    \sum_{\mathcal{V}_T} n_{\mathcal{V}_T} J_-\left(\frac{m_{\mathcal{V}_T}^2}{T^2}\right)
    +\sum_{\mathcal{V}_L} n_{\mathcal{V}_L}  J_-\left(\frac{m_{\mathcal{V}_L}^2}{T^2}\right) \right] \nonumber\\
    &-\frac{T^4}{2\pi^2}\sum_{\mathcal{V}_L} \frac{\pi}{6} \left(\frac{\overline{m}^3_{\mathcal{V}_L}}{T^3}-\frac{m_{\mathcal{V}_L}^3}{T^3}\right) \,,
    \label{eq:VT}
\end{align}
%-------
where the sum extends over fermions $f$ and bosons. The bosonic sector can be further divided into two categories: the transverse modes of gauge bosons, represented by $\mathcal{V}_T=W_T,Z_T$, and the longitudinal modes of gauge bosons and scalars, denoted by $\mathcal{V}_L=W_L,Z_L,\gamma_L,\Phi^0,\Phi^{\pm}$. The resummation of the $n=0$ Matsubara modes of the longitudinal components $\mathcal{V}_L$ leads to thermal corrections in their masses~\cite{Matsubara:1955ws,Quiros:1999jp}. The second line in~\autoref{eq:VT} corresponds to the Daisy contributions, where $\overline{m}_{\mathcal{V}_L}$ represents the thermal Debye mass calculated using the Arnold-Espinosa scheme~\cite{Arnold:1992rz,Basler:2016obg}. {The formulas for Debye masses are provided in Appendix~\ref{app:Debye}.}
Finally, the thermal functions for fermions $(J_+)$ and bosons $(J_-)$ are given by
%-------
\begin{align}
    J_{\pm}(x)=\mp\int_0^\infty dy~ y^2 \log\left(1\pm e^{-\sqrt{y^2+x^2}} \right)\,.
\end{align}
%-------

Whereas the effective potential in the electroweak phase transition is subject to theoretical uncertainties stemming from gauge parameter choices~\cite{Patel:2011th,Wainwright:2011qy,Metaxas:1995ab,Garny:2012cg,Chiang:2017nmu,Arunasalam:2021zrs,Hirvonen:2021zej,Lofgren:2021ogg}, Nielsen identities offer a way to construct gauge-independent probes~\cite{Nielsen:1975fs}. These identities ensure that the gauge dependence cancels out at the extrema of the potential
%--------
\begin{align}
    \frac{\partial V_\text{eff}(\Phi _1,\Phi _2,\xi )}{\partial \xi }=-C_i(\Phi _1,\Phi _2,\xi)\frac{\partial V_\text{eff}(\Phi _1,\Phi _2,\xi )}{\partial \phi_i }\,,
\end{align}
%-------
where $\xi$ is the gauge fixing parameter.  Inspired by the gauge independence guaranteed by Nielsen identities, we employ two distinct methods for phenomenological analyses. The first approach involves calculating the finite-temperature effective potential and performing a numerical scan. The second approach focuses on determining the gauge-invariant vacuum upliftment at $T=0$. In~\autoref{sec:upliftment}, we highlight that the upliftment of the true vacuum relative to the symmetric vacuum at zero temperature serves as an effective probe of the phase transition's strength.
While the first method carries uncertainties associated with gauge parameter choices, the latter approach is gauge invariant, as assured by Nielsen identities~\cite{Dorsch:2017nza,Patel:2011th}. It is worth noting that we introduce additional counterterms at one-loop order to preserve the positions of the electroweak vacuum and masses. The agreement between our numerical scan and the profile derived from the vacuum upliftment serves to confirm the reliability of the numerical scan despite its inherent uncertainties.\footnote{{For a comprehensive analysis of the uncertainties associated with finite temperature effective potential, we refer to~\cite{Athron:2022jyi}. }}

%%%%%%%%%%%%%%%%%%%%%%%%%%%%%%%%%%%%%%%%%%%%%%%%%%%%%%%%%%%%%%%%%%%%%%%%%%%%%%
\section{Electroweak Phase Transition and Gravitational Waves}
\label{sec:EWPT_GW}
%%%%%%%%%%%%%%%%%%%%%%%%%%%%%%%%%%%%%%%%%%%%%%%%%%%%%%%%%%%%%%%%%%%%%%%%%%%%%%

The finite temperature effective potential dictates the phase-transition pattern. The two Higgs doublet model displays both single and multi-step phase transitions. The first-order phase transition occurs through tunneling from false  to true vacua. It results in bubbles of the broken phase that pop up and expand in the surrounding region of the symmetric phase, transitioning from the false vacuum to the true vacuum. The tunneling probability is given by ~\cite{Linde:1980tt,Coleman:1977py}
%-----
\begin{align}
    \Gamma (T)\approx T^4\left (\frac {S_3}{2\pi T}  \right )^{3/2}e^{-\frac {S_3}{T}}\,,
\end{align}
%-----
where $S_3$ represents the three-dimensional Euclidean action associated with the critical bubble formation
%-----
\begin{align}
    S_3=4\pi\int_{0}^{\infty}{dr r^2\left [ \frac 12\left ( \frac {d\phi(r)}{dr} \right )^2+V(\phi,T) \right ]}\,.
\end{align}
%-----
Here, the scalar field $\phi$ corresponds to the critical bubble profile, which is determined by solving the following differential equation
%-----
\begin{align}
    \frac {d^2\phi}{dr^2}+\frac 2r \frac{d\phi}{dr}=\frac {dV(\phi,T)}{d\phi}\,, \quad   \text{with} \quad \lim_{r\rightarrow \infty}\phi(r)=0
    \quad \text{and} \quad \lim_{r\rightarrow 0}\frac {d\phi(r)}{dr}=0.
    \label{eq:tunneling}
\end{align}
%-----
We utilize the publicly available code {\tt CosmoTransitions}~\cite{Wainwright:2011kj} to solve the differential equation and compute the Euclidean action $S_3$.

The first-order phase transition is considered to be completed around the nucleation temperature $T_n$, which corresponds to the point where one bubble nucleates per unit horizon volume~\cite{Moreno:1998bq}.
%-----
\begin{align}
    \int_{T_n}^{\infty}\frac {dT}{T}\frac {\Gamma (T)}{H(T)^4}=1\,.
\end{align}
%-----
This condition ensures that the bubbles percolate even in the inflating Universe. For the electroweak phase transition, with a nucleation temperature of approximately $T_n \approx 100~\mathrm{ GeV}$, this condition can be approximated as~\cite{Quiros:1999jp}
%-----
\begin{align}
\frac {S_3(T)}{T}\approx 140\,.
\label{eq:Tncond}
\end{align}
%-----

To preserve the baryon asymmetry generated through electroweak baryogenesis, it is crucial to suppress the sphaleron  process inside the bubble. This requires the electroweak symmetry breaking to undergo a strong first-order phase transition~\cite{Quiros:1999jp}
%-----
\begin{align}
   \xi_c\equiv \frac{v_c}{T_c} \gtrsim 1\,,
   \label{eq:xic}
\end{align}
%-----
where {$v_c\equiv\sqrt{\omega_1^2(T_c)+\omega_2^2(T_c)+\omega_{CB}^2(T_c)}$}
% $v_c\equiv \sqrt{\omega_1^2(T_c)+\omega_2^2(T_c)+\omega_{CP}^2(T_c)+\omega_{CB}^2(T_c)} $
is the Higgs VEV at the critical temperature $T_c$. This critical temperature corresponds to the point where the broken and unbroken vacua of the electroweak symmetry are degenerate. The approximate inequality in~\autoref{eq:xic} indicates the theoretical uncertainty in this condition~\cite{Patel:2011th}.

The production of stochastic gravitational waves is a significant consequence of a first-order phase transition. These GW originate from three main sources: the collision of vacuum bubbles, fluid motion resembling sound waves in the plasma, and turbulent motion within the plasma. Each source contributes to the GW spectrum, which can be described by numerical functions dependent on two parameters that capture the dynamics of the phase transition at the nucleation temperature $T_n$~\cite{Grojean:2006bp,Caprini:2015zlo,Alves:2019igs,Alves:2020bpi}.\footnote{Comprehensive information on the computation of gravitational wave signals can be found in Appendix~\ref{app:GW}.} The first parameter is $\alpha$, defined as the ratio of the latent heat released during the phase transition ($\epsilon$) to the energy density of the vacuum radiation ($\rho_{rad}$), \emph{i.e.}, $\alpha\equiv \epsilon/\rho_{rad}$. The latent heat  and the vacuum radiation energy density are expressed as
%-----
\begin{align}
 \epsilon = \Delta\left(- V_{\rm eff} + T\frac{\partial V_{\rm eff}}{\partial T}\right)_{T=T_n} \quad \text{and}\quad \rho_{\rm rad} = \frac{\pi^2}{30}g_\star T_n^4\,,
\end{align}
%-----
where $\Delta$ represents the difference between the true and false vacua, and
$g_\star$  the number of relativistic degrees of freedom in the plasma.
The second important parameter is $\beta/H_n$, which characterizes the inverse time duration of the phase transition. This quantity is defined as\footnote{{ The derivative $\frac{d}{dT}\left(\frac{S_3}{T}\right)$ is calculated using the fourth-order finite difference formula with $\Delta T=0.01~\mathrm{GeV}$.}}
%-----
\begin{align}
	\frac{\beta}{H_n} &\equiv T_n\frac{d}{dT}\left.\left(\frac{S_3}{T}\right)\right|_{T=T_n}\,,
\end{align}
%-----
where $H_n$ is the Hubble constant at the nucleation temperature $T_n$. Detectable GW signals are typically associated with a slow phase transition (small $\beta/H_n$) and a large latent heat release (large $\alpha$).

Finally, to assess the detectability of GW signal, we employ the signal-to-noise ratio (SNR) measure~\cite{Caprini:2015zlo}
%-----
\begin{align}
\mathrm{SNR}=\sqrt{\mathcal{T} \int_{f_{\min }}^{f_{\max }} d f\left[\frac{h^{2} \Omega_{\mathrm{GW}}(f)}{h^{2} \Omega_{\mathrm{Sens}}(f)}\right]^{2}}\,,
\end{align}
%-----
where $\Omega_{\rm Sens}$ represents the sensitivity curve of the considered GW detector~\cite{LISA:2017pwj} and $\mathcal{T}$ corresponds to the mission duration. For our analysis, we adopt the LISA gravitational wave detector as a benchmark, with $\mathcal{T}=5$ years and a detection threshold of ${\rm SNR}=10$~\cite{Caprini:2015zlo}.

{The existence of percolation temperature $T_p$, where $29\%$ of space is covered by bubbles, guarantees the completion of phase transition~\cite{Athron:2022mmm}. Supercooling emerges when the nucleation temperature is substantially lower than the critical temperature, leading to a pronounced $\alpha \gg 1$~\cite{Badger:2022nwo,Athron:2022mmm}. In our analysis, the overwhelming majority of data points in our scan, specifically $99.99\%$, display $\alpha<1$. Hence, we assume there is no supercooling and the percolation temperature can be approximated to the nucleation temperature $T_p\simeq  T_n$.
}

%%%%%%%%%%%%%%%%%%%%%%%%%%%%%%%%%%%%%%%%%%%%%%%%%%%%%%%%%%%
\section{Barrier Formation and Vacuum Upliftment}
\label{sec:upliftment}
%%%%%%%%%%%%%%%%%%%%%%%%%%%%%%%%%%%%%%%%%%%%%%%%%%%%%%%%%%%
Introducing a second Higgs doublet to the SM Higgs sector can alter the behavior of electroweak symmetry breaking from a smooth crossover to a strong first-order phase transition. In Ref.~\cite{Goncalves:2021egx}, the authors studied the key ingredients that trigger this transmutation in the EWPT by focusing on the barrier formation and upliftment of the true vacuum in the context of CP-conserving 2HDM~\cite{Goncalves:2021egx, EWPT-NMSSM, EWPT-Nature, Dorsch:2017nza}. In this model, the barrier is driven primarily by one-loop corrections and $\xi_c$ can be correlated with $ \Delta \mathcal{F}_0/|\mathcal{F}_0^{\rm SM}|$, a gauge independent parameter calculated at zero temperature. The $  {\Delta \mathcal{F}_0}/{|\mathcal{F}_0^{\rm SM}|}$ is defined as
 %-----
\begin{align}
    \frac {\Delta \mathcal{F}_0}{|\mathcal{F}_0^{\rm SM}|} \equiv \frac {\mathcal {F}_0-\mathcal{F}_0^{\rm SM}}{|\mathcal{F}_0^{\rm SM}|},
\end{align}
%-----
where $\mathcal{F}_0$ is the zero-temperature vacuum energy density of the 2HDM defined as
%-----
\begin{align}
    \mathcal{F}_0\equiv V_{\rm eff}(v_1,v_2,T=0)-V_{\rm eff}(0,0,T=0),
\end{align}
%-----
with $\mathcal{F}_0^{\rm SM}=-1.25 \times 10^8 ~\text{GeV}^4$.
%----
\begin{figure}[!t]
\includegraphics[width=0.49\textwidth]{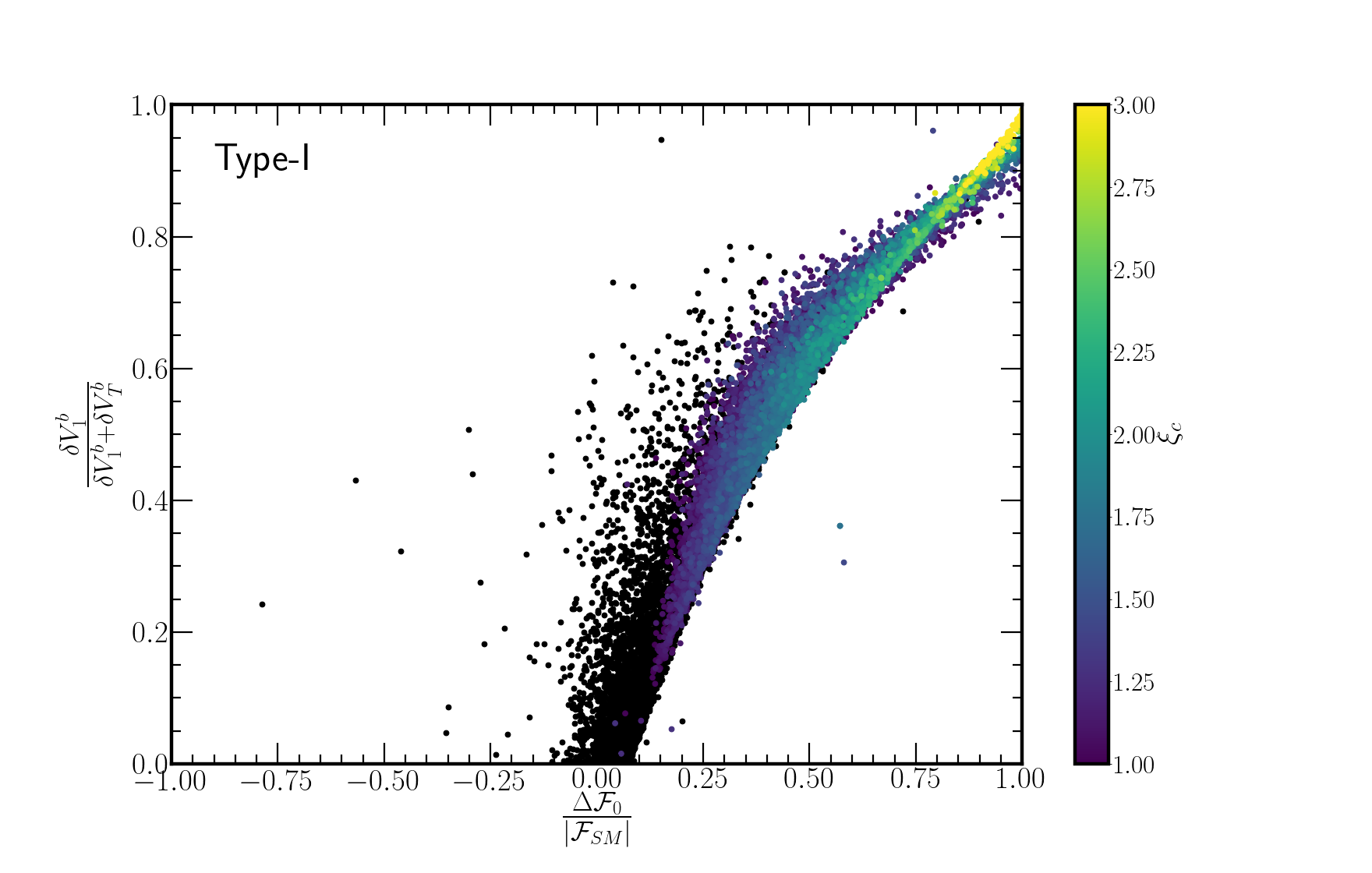}
\includegraphics[width=0.49\textwidth]{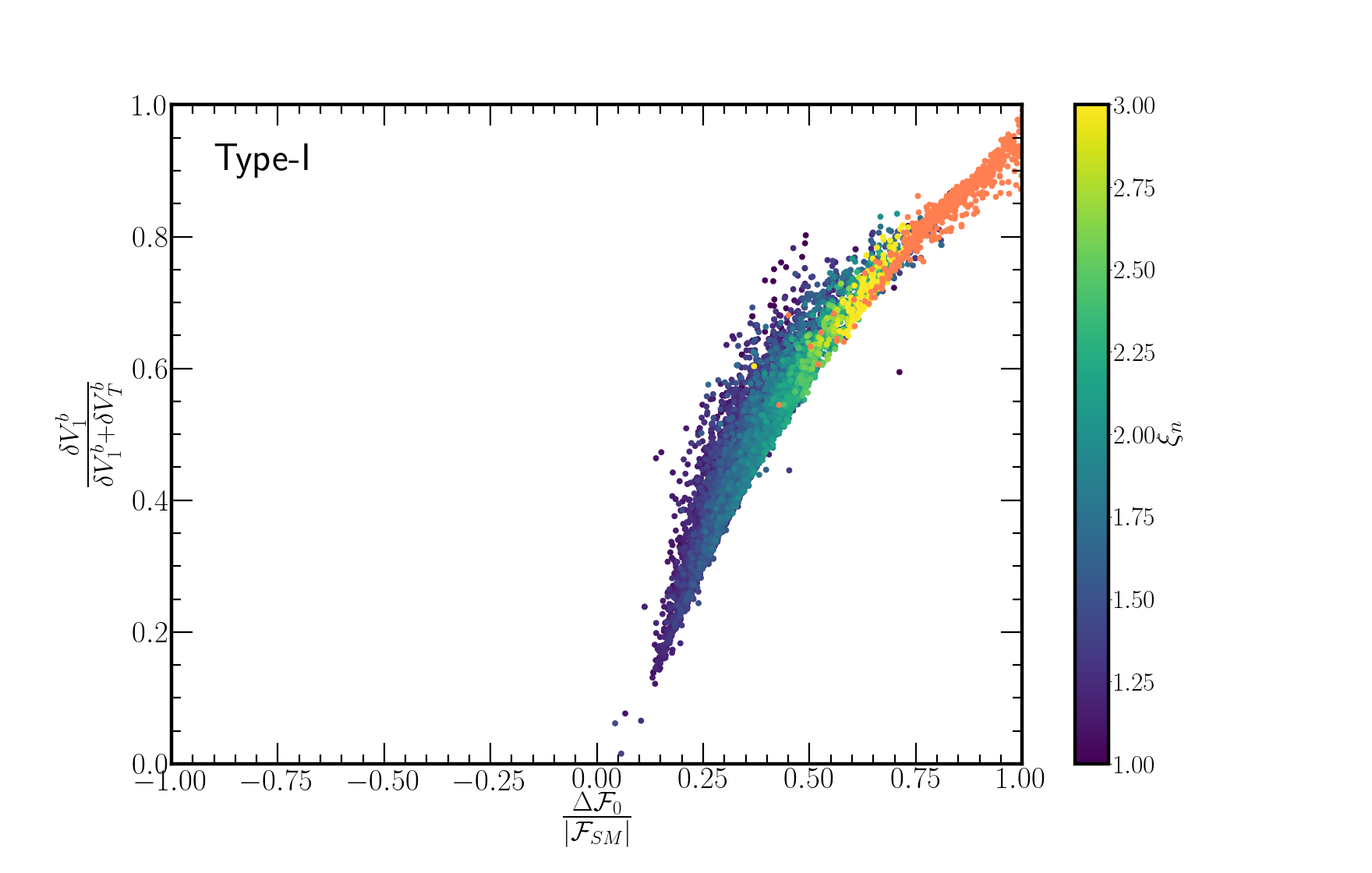}
\caption{The ratio $\frac {\delta V_1^b}{\delta V_1^b+\delta V_T^b}$ for the barrier at $T_c$ versus $\Delta \mathcal{F}_0/\mathcal{F}_0^{\rm SM}$ color coded with $\xi_c$. Black denotes all first-order phase transition points with  $0<\xi_c<1$. In the right panel, we have the very same $\frac{\delta V_1^b}{\delta V_1^b+\delta V_T^b}$ for the barrier at $T_c$ versus $\Delta \mathcal{F}_0/\mathcal{F}_0^{\rm SM}$ color coded with $\xi_n$ for the parameter points that have a solution to \autoref{eq:Tncond}, thereby they have a nucleation temperature. The orange points represent parameter space configurations where the Universe is trapped in the false vacuum and the phase transition is incomplete. In these plots, we only considered the parameter points where the barrier is generated by one-loop and thermal corrections $\delta V_1^b,\delta V_T^b>0$, which covers $99\%$ of the parameter space points.}
\label{fig:dF0-xic}
\end{figure}
%----

%----
\begin{figure}[t!]
\includegraphics[width=0.49\textwidth]{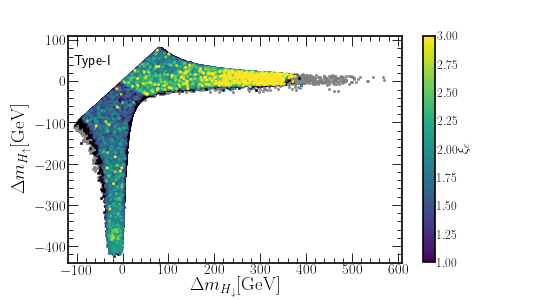}
\includegraphics[width=0.49\textwidth]{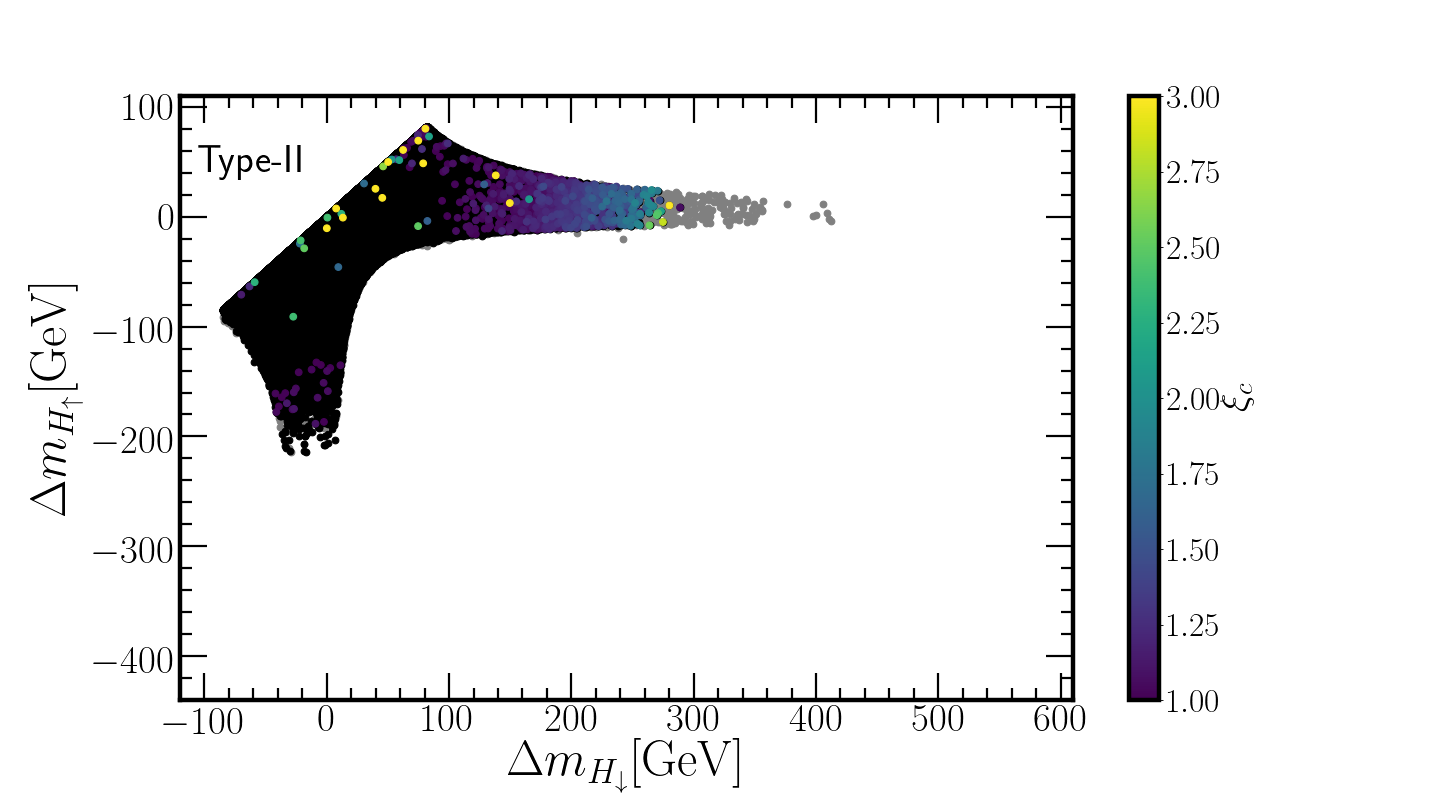}\\
\includegraphics[width=0.49\textwidth]{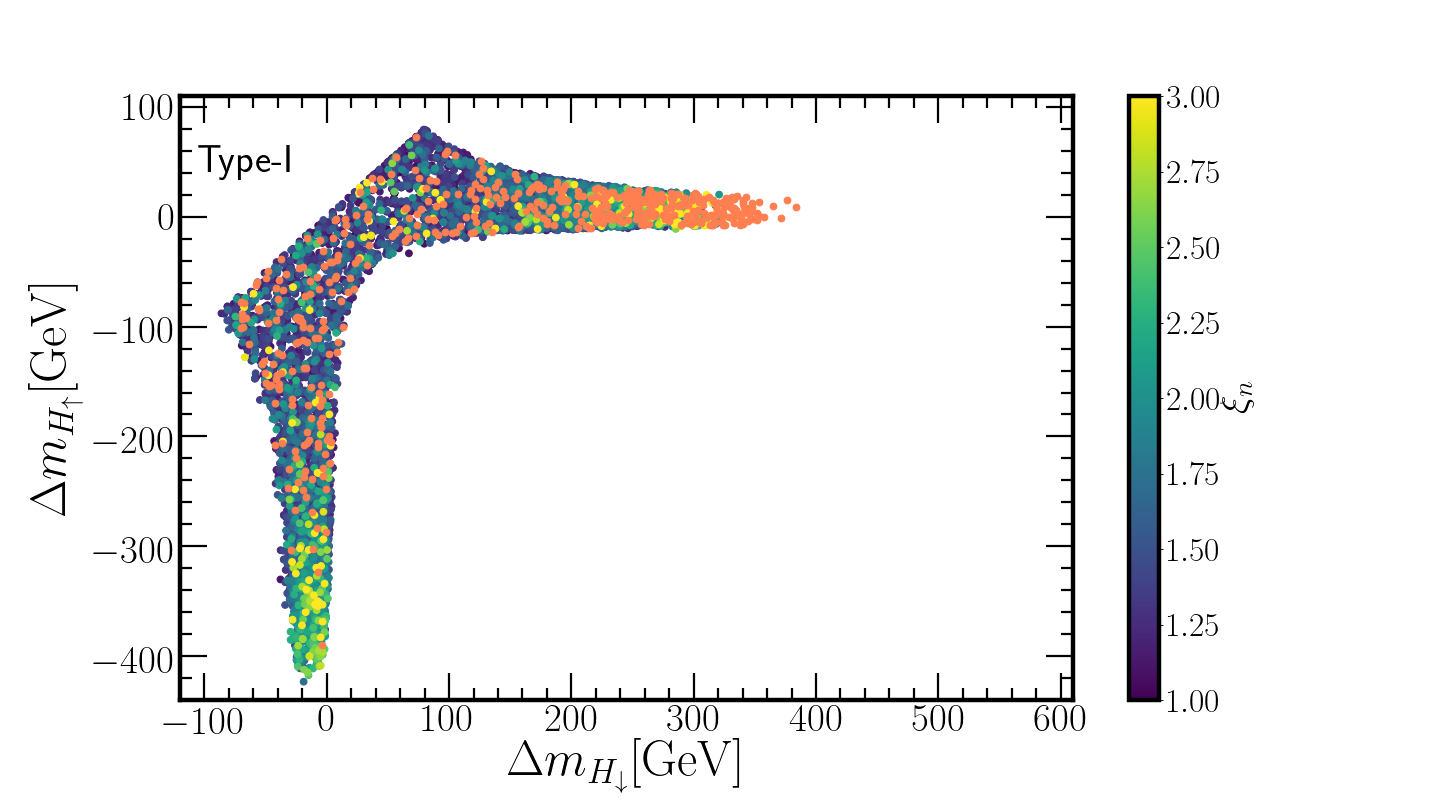}
\includegraphics[width=0.49\textwidth]{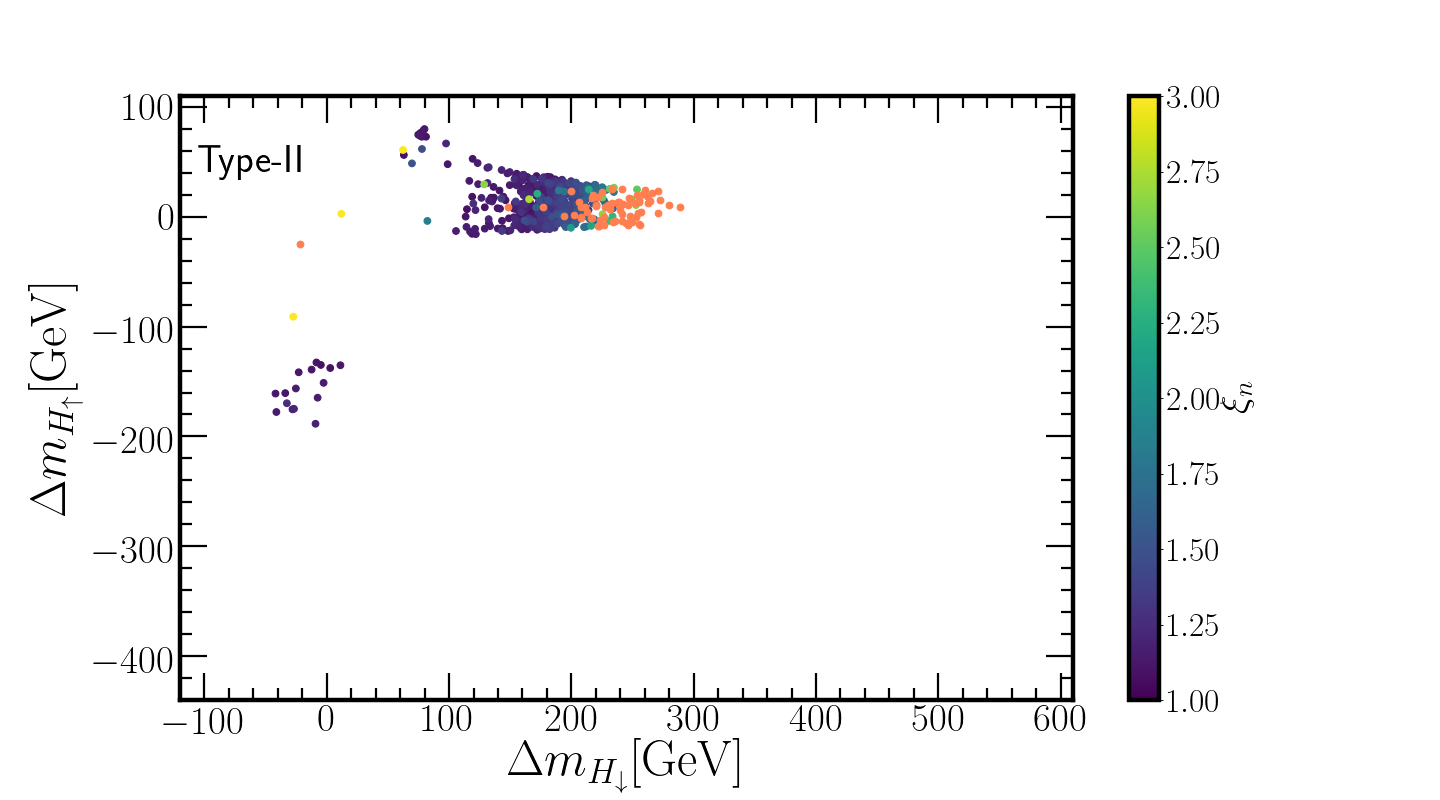}\\
\includegraphics[width=0.49\textwidth]{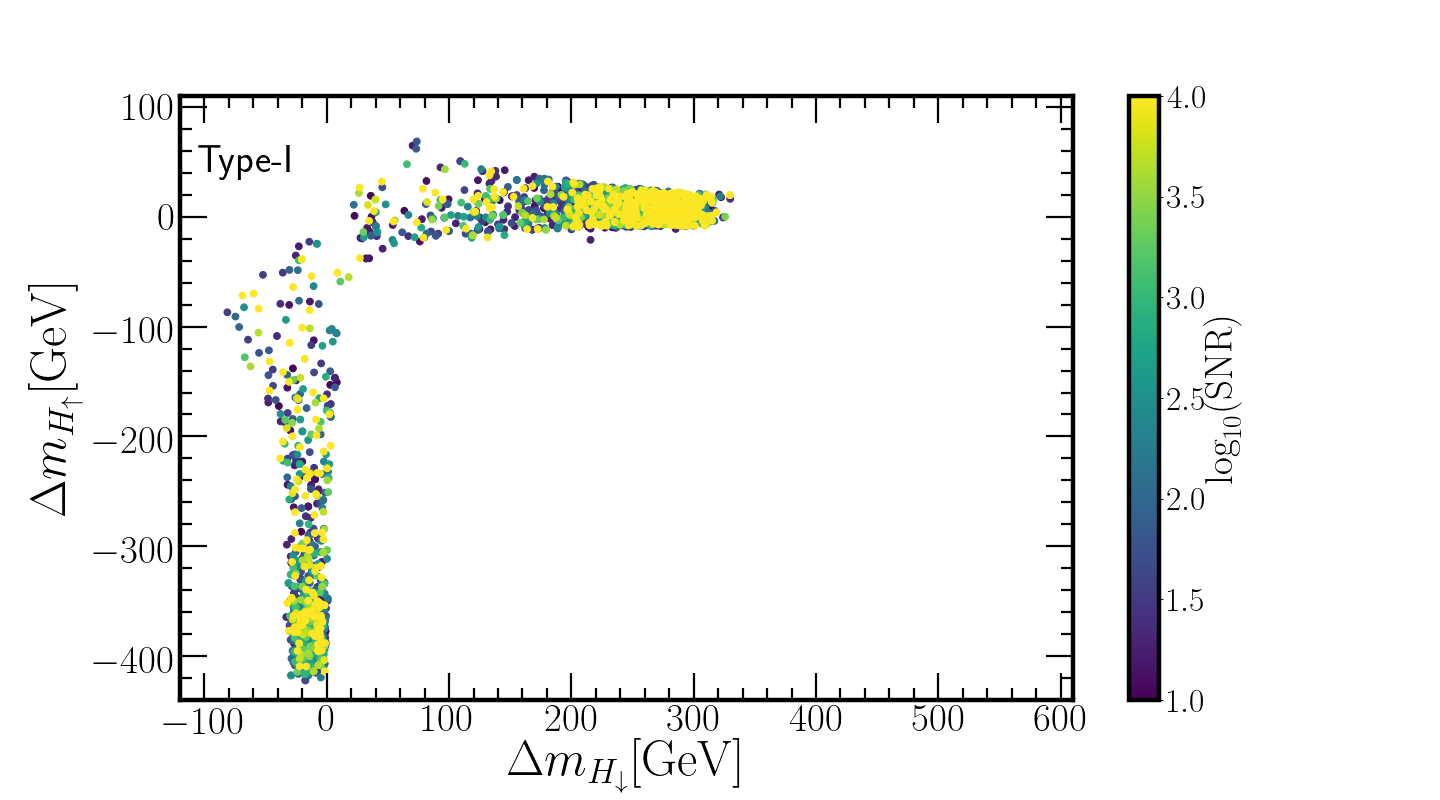}
\includegraphics[width=0.49\textwidth]{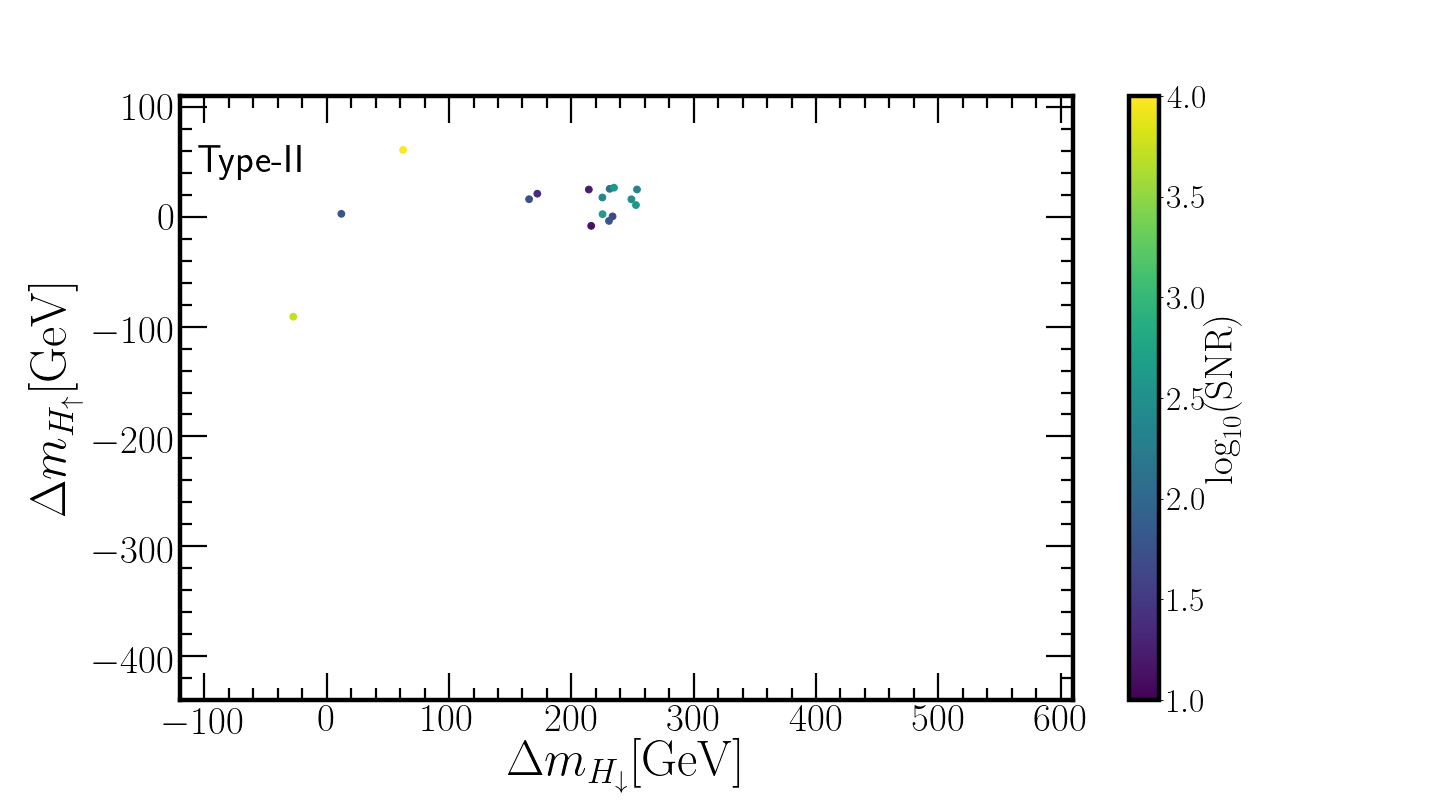}
\caption{The parameter space scan in terms of ($\Delta m_{H_\uparrow}$, $\Delta m_{H_\downarrow}$) for Type-I (left panel) and Type-II (right panel). The heat map tracks order parameter $\xi_c$ (upper panel), $\xi_n$ (middle panel), and $\log_{10}(\mathrm{SNR})$ (lower panel). The gray points in the upper panel pass all theoretical and current experimental constraints. Black points also show first-order phase transition with $0<\xi_c<1$.  {The samples with $\xi_c<1$ have been excluded from the middle and bottom panels.} In the middle panel, orange color represents parameter points trapped in false vacuum and phase transition is incomplete.  The parameter space scan is implemented using {\tt ScannerS}~\cite{Muhlleitner:2020wwk}, where we impose the constraints from perturbative unitarity, boundedness from below, vacuum stability, electroweak precision,  flavor constraints, and EDM limits. {\tt HiggsBounds} and {\tt HiggsSignals} are used to incorporate the searches for additional scalars as well as the 125~GeV Higgs boson measurements~\cite{Bechtle:2020pkv,Bechtle:2020uwn}.}
\label{dmhp_dmhl}
\end{figure}
%----

It is interesting to examine whether the phase transition features of the CP-conserving 2HDM prevail in the CP-violating 2HDM.  In~\autoref{fig:dF0-xic} (left panel), we show that the fraction of one-loop contribution to the barrier height is correlated with the zero temperature vacuum upliftment measure $\Delta \mathcal{F}_0/ | \mathcal{F}_0^{\rm SM}|$. We observe that the larger the one-loop correction, the higher the value of vacuum upliftment. In particular, this correlation can be seen for $\xi_c\gtrsim 1$. As one-loop effects are the dominant contributions, we can use $\Delta \mathcal{F}_0/ | \mathcal{F}_0^{\rm SM}|$ to shed light on the properties of the EWPT. We can approximately propose $\Delta \mathcal{F}_0/ | \mathcal{F}_0^{\rm SM}| \gtrsim 0.2$ as minimal condition for strongly first-order EWPT in the CP-violating 2HDM.

In the scenario where the vacuum upliftment measure $\Delta \mathcal{F}_0/ | \mathcal{F}_0^{\rm SM}|$ is extremely large, the tunneling from false vacuum to true vacuum becomes challenging, translating into \autoref{eq:Tncond} having no solution~\cite{Goncalves:2021egx, Biekotter:2022kgf}. Thus, the Universe is trapped in a high energetic electroweak symmetric vacuum, yielding a nonphysical vacuum. In the \autoref{fig:dF0-xic} (right panel), we denote these points with the orange color. Most of the parameter points with vacuum upliftment measure $\Delta \mathcal{F}_0/ | \mathcal{F}_0^{\rm SM}| \gtrsim 0.87$ exhibit a vacuum trapped scenario. The above constraint excludes the bulk of $\xi_c>2$ points, which would otherwise serve as promising candidates for successful electroweak baryogenesis.

In~\autoref{dmhp_dmhl} (upper-panel), we show the scanned points in the $(\Delta m_{H_\uparrow}$, $\Delta m_{H_\downarrow})$ plane color-coded by $\xi_c$, where $m_{H_\uparrow}$ ($m_{H_\downarrow}$) represents the mass of the heaviest (lightest) BSM neutral scalar, and
$\Delta m_{H_\uparrow} \equiv m_{H^\pm}-m_{H_\uparrow} $  ($\Delta m_{H_\downarrow} \equiv m_{H^\pm}-m_{H_\downarrow} $). The gray points in the background pass all the theoretical and current experimental constraints. The black points also satisfy the first-order phase transition condition with $0<\xi_c<1$. The preference to the region with  $m_{H_\uparrow} \approx m_{H^\pm}$ or $m_{H_\downarrow} \approx m_{H^\pm}$ is induced predominantly by electroweak precision measurements~\cite{Grimus:2007if,Gerard:2007kn}.  The $\xi_c>1$ points favor a large value of  $|\Delta m_{H_\uparrow}|$ or $|\Delta m_{H_\downarrow}|$ because a higher value of $\Delta \mathcal{F}_0/ | \mathcal{F}_0^{\rm SM}|$ requires a larger mass split, similarly to the CP-conserving scenario~\cite{Goncalves:2021egx}. In~\autoref{dmhp_dmhl} (middle-panel), we present the parameter points in the $(\Delta m_{H_\uparrow}, \Delta m_{H_\downarrow})$ plane, color-coded by $\xi_n$. The points marked in orange correspond to locations where vacuum trapping occurs. The majority of parameter points with large values of $\xi_c$, where $m_{H_\uparrow} \approx m_{H^\pm}$ and $m_{H^\pm}-m_{H_\downarrow}>250~\mathrm{GeV}$, are trapped in the false vacuum state. As a result, the phase transition remains incomplete.
%-----------------------------
 \begin{figure}[t!]
    \centering
    \includegraphics[width=0.7\hsize]{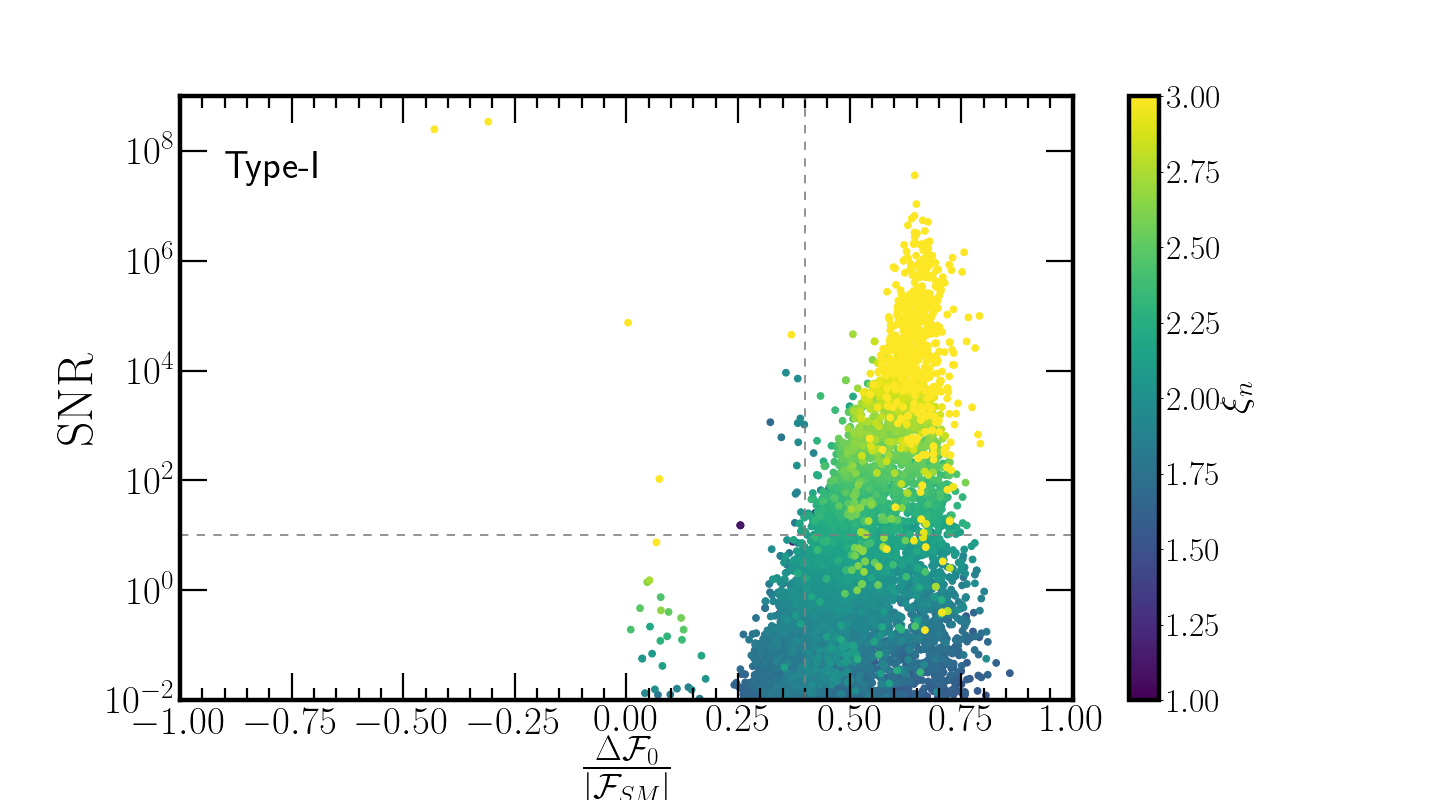}
    \caption{SNR versus $\Delta \mathcal{F}_0/\mathcal{F}_0^{\rm SM}$ color coded with $\xi_n$ for the Type I parameter point with $\xi_n>1$. The dotted line in the plot corresponds to an SNR value of 10, serving as the threshold above which LISA can probe the parameter points.}
    \label{fig:dF0SNR}
\end{figure}
%------------------------------

In ~\autoref{dmhp_dmhl} (lower-panel), we show parameter points that can be probed by LISA in the $(\Delta m_{H_\uparrow}$, $\Delta m_{H_\downarrow})$ plane. The color coding in this case represents the logarithm (base 10) of the signal-to-noise ratio. We focus on points above the SNR threshold, $\rm SNR>10$. Among these points, we highlight the benchmark point BP4 in \autoref{tab:1}, which serves as an example of a parameter point that can be probed by LISA. For the parameter points with $\xi_n>1$, $6\%$ of Type-I points  show a detectable GW signal by LISA, whereas it is around $2.5\%$ for Type-II. These differences between type-I and type-II scenarios are driven by constraints from flavor physics~\cite{Goncalves:2021egx}. More concretely, constraints from $B$-meson decays impose a lower bound on the charged scalar mass requiring $m_{H^\pm}\gtrsim 580$~GeV in the type-II 2HDM.  In~\autoref{fig:dF0SNR}, we show the correlation between SNR and zero-temperature vacuum upliftment measure $\Delta \mathcal{F}_0/\mathcal{F}_0^{\rm SM}$ for the Type I parameter points with $\xi_n>1$. The bulk of parameter points that exhibit strong GW signals are associated with large $\Delta \mathcal{F}_0/\mathcal{F}_0^{\rm SM}$ measure. In most cases, parameter points with $\Delta \mathcal{F}_0/\mathcal{F}_0^{\rm SM}<0.4$ do not show a promissable GW at LISA.

%----
\begin{table}
\centering
\begin{tabular}{ c p{3.0cm}  p{2.5cm} p{2.5cm} p{2.5cm} p{2.5cm}  }
\toprule[1pt]
\toprule[1pt]
& BP1  &BP2 & BP3& BP4  \\
\cmidrule[0.5pt]{2-5}
$m
_{H_\downarrow}$ [GeV]& $178.596$& $227.340$&  $237.815$&$151.408$\\
$m_{H_\uparrow}$ [GeV]& $348.814$& $446.822$&  $440.556$&$443.706$ \\
$m_{H^+}$ [GeV]& $368.266$& $222.252$& $219.544$& $438.803$ \\
$t_\beta$& $ 15.069$& $11.346$& $16.622$& $ 11.316$\\
$m_{12}^2$ [GeV$^2$]& $4471.227$& $23658.001$& $3837.544$&$2043.461$\\
$\alpha_c$& $-0.1816$& $-0.0030$& $-0.2404$& $ -0.0410$\\
$\cos(\beta-\alpha)$ & $0.2628$& $ 0.2129$&  $0.1544$& $0.1316$\\
\hline
$\xi_c$ & $1.575$& $1.161$& $1.519$& $2.13$\\
$T_c$ [GeV]& $128.65$& $127.54$& $120.24$& $106.65$\\
$T_n$ [GeV]& $121.04$& $127.22$& $ 116.81$& $80.78$  \\
$\alpha$& $ 0.0227$& $0.0142$& $0.0262$& $0.0642$\\
$\beta/H_n$& $2.488\times 10^3$& $1.490\times 10^6$& $7.186\times 10^3$& $4.744\times10^2$\\
SNR & $ 2.73\times 10^{-1}$& $3.01 \times 10^{-13}$& $9.72\times 10^{-3}$& $1.09\times 10^4$ \\
$\eta_B^{\mathrm{tun}}$& $5.51\times 10^{-12}$& $1.09\times 10^{-10}$& $1.16\times 10^{-12}$& $1.43 \times 10^{-13}$\\
$\eta_B^{\mathrm{Kink}}$& $1.06\times 10^{-11}$& $8.79\times 10^{-14}$& $-2.33\times 10^{-12}$& $4.10\times 10^{-13}$\\
\bottomrule[1pt]
\bottomrule[1pt]
\end{tabular}
\caption{Benchmark points for the complex two Higgs double model.}
\label{tab:1}
\end{table}
%----

%%%%%%%%%%%%%%%%%%%%%%%%%%%%%%%%%%%%%%%%%%%%%%%%%%%%%%%%%%%
\section{Baryon Asymmetry Calculation}
\label{sec:BAU}
%%%%%%%%%%%%%%%%%%%%%%%%%%%%%%%%%%%%%%%%%%%%%%%%%%%%%%%%%%%
\subsection{Estimation of bubble wall profile}
\label{subsec:tunprof}
%%%%%%%%%%%%%%%%%%%%%%%%%%%%%%%%%%%%%%%%%%%%%%%%%%%%%%%%%%%
The bubble profile in the radial coordinate can be obtained by solving the tunneling equation~\autoref{eq:tunneling}. The baryon asymmetry calculation is performed in the bubble wall coordinate system $z$, where $z=0$ denotes the bubble wall. We obtain the position of the bubble wall in the radial coordinate system $r_0$, where the energy density obtains the maximum value. The energy density $U_{E}$ is given by
%----
\begin{align}
U_E(r)=\frac12\left [\phi'(r) \right ]^2+V_{\mathrm{eff}}\left ( \phi(r) \right )\,,
\end{align}
%----
and the bubble wall coordinate $z$ can be defined as
%----
\begin{align}
z=r-r_0.
\label{eq:r0}
\end{align}
%----

\begin{figure}[t!]
\includegraphics[width=0.49\textwidth]{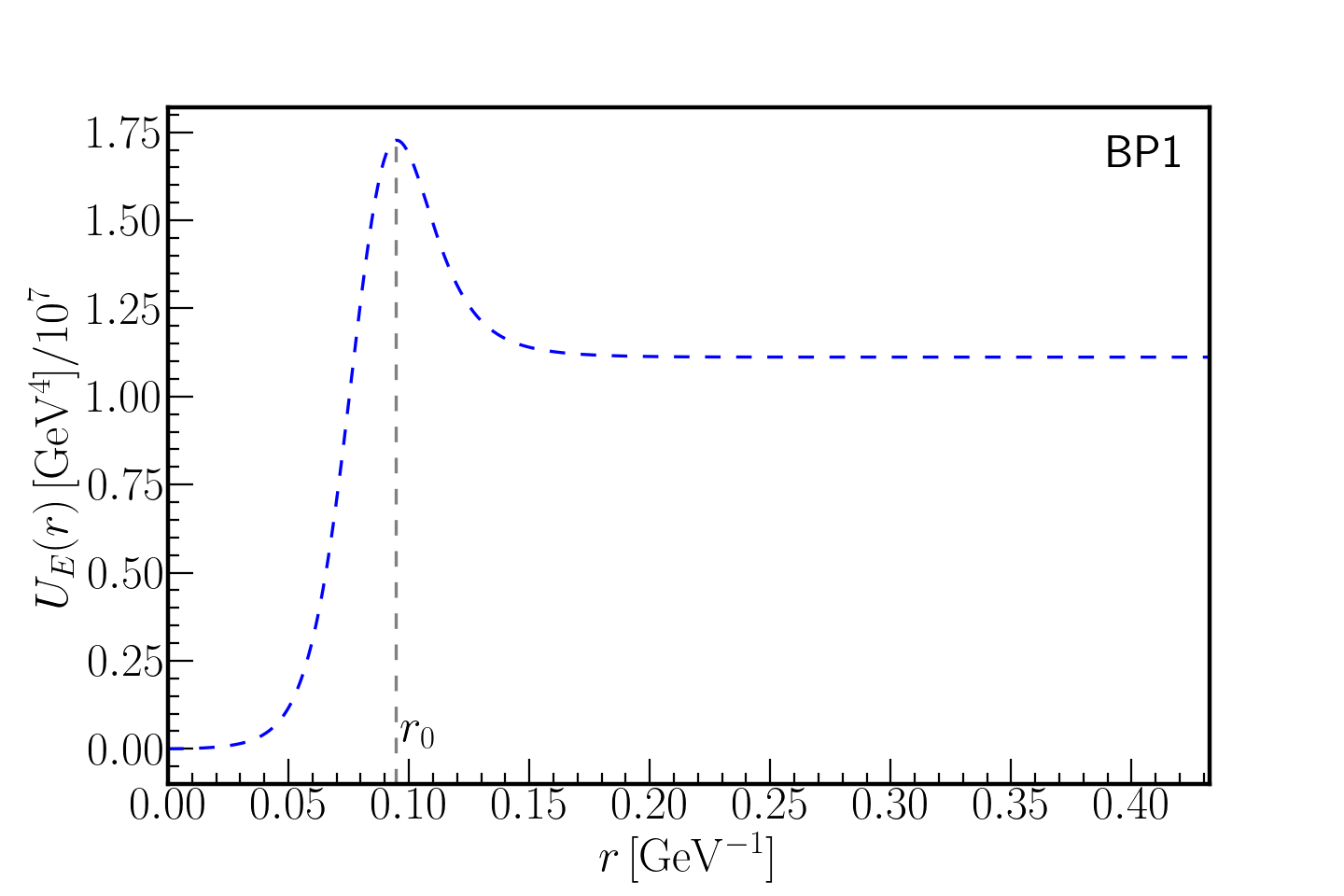}
\includegraphics[width=0.49\textwidth]{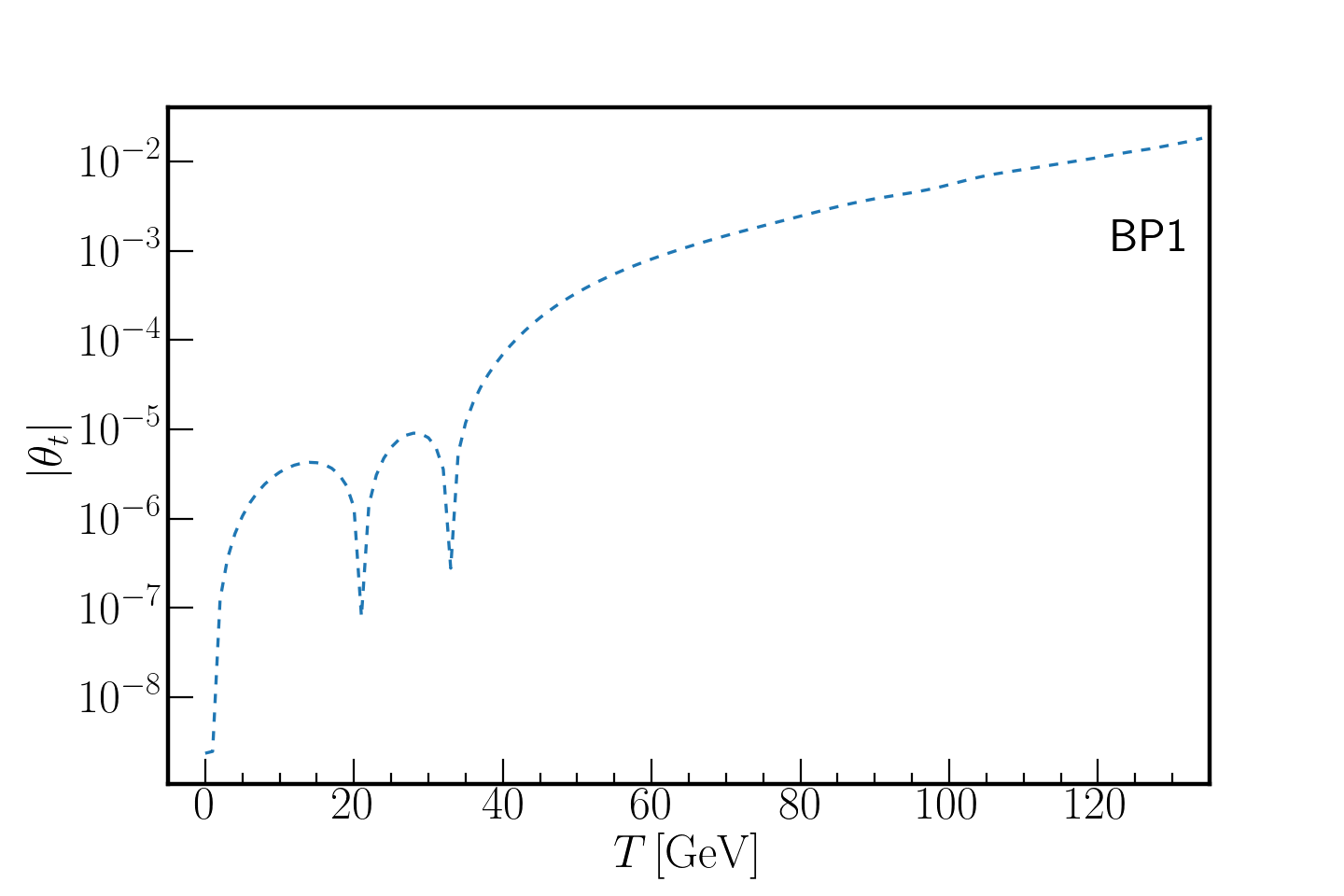}
\caption{The left panel shows energy density $U_E(r)$ as a function of $r$ for BP1 as defined in   \autoref{tab:1}. The position of the bubble wall $r_0$ is determined as maxima of $U_E(r)$. In the right panel, we show the dynamic evaluation of CP violating angle of top quark $\theta_t$ with temperature in the broken phase.}
\label{BP1UEtb1}
\end{figure}

A key ingredient for the baryogenesis is the complex mass of quarks and leptons, which couples to $\Phi_2$,
%----
\begin{align}
    m_i(z) = \frac{y_i}{\sqrt{2}}\omega_2e^{-i\omega_\theta} \equiv |m_i(z)|e^{i\theta^i(z)}
\end{align}
%----
To illustrate these concepts, we present a graphical representation in \autoref{BP1UEtb1}. The left panel displays the energy density $U_E(r)$ as a function of the radial distance $r$, specifically for benchmark point BP1 as defined in \autoref{tab:1}.  The position of the bubble wall is identified as the barrier of the tunneling profile. On the right panel, we show the dynamic variation of the CP-violating angle of the top quark with respect to temperature in the broken phase for the BP1. As thermal effects come into play, additional CP violation is induced at higher temperatures. This effect becomes prominent, whereas at zero temperature, it is roughly seven orders of magnitude smaller. The  oscillatory behavior observed  between temperatures of  $20$~GeV and $35$~GeV arises due to thermal contributions that lead to a change in sign of the CP angle, which we represent in terms of the absolute value $|\theta_t|$.

In the literature, it is a customary practice to parameterize the tunneling profile $\theta^i(z)$ by kink profile~\cite{Bodeker:2004ws,Fromme:2006wx, Fromme:2006cm,Basler:2020nrq, Basler:2021kgq}
%----
\begin{align}
    \theta^i(z)=\frac{\theta^i_{\mathrm{brk}}+\theta^i_{\mathrm{sym}}}{2}-\frac{\theta^i_{\mathrm{brk}}-\theta^i_{\mathrm{sym}}}{2}\tanh\left ( \frac{z}{L_W} \right )  ,
\end{align}
%----
where $\theta_\text{brk}^i$ ($\theta_\text{sym}^i$) is the phase at the broken (symmetric) minimum.  The thickness of the wall $L_W$ is given by $L_W=v_n/\sqrt{8 V_b}$~\cite{Fromme:2006wx} with $v_n$ representing the VEV  at EWPT and $V_b$  the height of the barrier that separates the two  minima (at the nucleation temperature $T_n$).
Remarkably, this parameterization displays a tunneling profile that is symmetric with respect to the bubble wall. In Section~\ref{sec:BassymC2HDM}, we compare the estimation of the baryon asymmetry of the universe using two different methods: the kink profile and the explicit solution from the bubble profile. In the latter case, the bubble profile is obtained by directly solving the tunneling equation. By comparing the results obtained from these two approaches, we can evaluate the consistency and reliability of the BAU estimation.

%----
%----

%%%%%%%%%%%%%%%%%%%%%%%%%%%%%%%%%%%%%%%%%%%%%%%%%%%%%%%%%%%
\subsection{Semi-classical force method}
\label{subsec:FHmethod}
%%%%%%%%%%%%%%%%%%%%%%%%%%%%%%%%%%%%%%%%%%%%%%%%%%%%%%%%%%%

The baryon asymmetry in the Universe can be estimated using the semi-classical force method. This framework utilizes the existence of a fermion with varying complex mass as it passes through the bubble wall. The particle interaction with the bubble wall can be formalized using the WKB approximation~\cite{Fromme:2006wx, Fromme:2006cm, Cline:2020jre} or the closed-time-path formalism of thermal field theory~\cite{Kainulainen:2001cn, Kainulainen:2002th, Prokopec:2004ic, Prokopec:2003pj}, where the force acting on the particle is given by
%----
\begin{align}
F_z=-\frac{(m^2)^\prime}{2E_0}\pm s \frac{(m^2\theta^\prime)^\prime}{2E_0E_{0z}}\mp\frac{\theta^\prime m^2(m^2)^\prime}{4E_0^3E_{0z}}.
\label{eq:force}
\end{align}
%----
The radial coordinate denotes the perpendicular distance from the wall in the rest frame of the wall, where the positive direction of $z$ points towards the symmetric phase. $E_0$ is the conserved wall frame energy of the quasi-particle, $E_{0z}^2=E_0^2-p_{\parallel}^2$ and $(..)^\prime$ denotes the derivative with respect to the $z$ coordinate. The first term in \autoref{eq:force} conserves CP, whereas the second and third terms depend on the spin and nature of the particle, with the upper sign solution corresponding to the particle and the lower sign to the antiparticle. Thus, the presence of a non-zero value for $\theta^\prime$ generally indicates the appearance of CP violation~\cite{Fromme:2006cm, Fromme:2006wx,Kainulainen:2002th}. Assuming that the kinetic momentum is conserved in collisions, the perturbation $\delta f_i$ from the equilibrium density $f_i$ of species $i$ caused by the movement of the bubble wall is given by
%----
\begin{align}
f_i=\frac {1}{e^{\beta\left [ \gamma_W(E_0+v_wp_z)-\mu_i \right ]}\pm1}+\delta f_i,
\label{eq:distribution}
\end{align}
%----
where $\beta=1/T$, $\gamma_W= 1/\sqrt{1-v_w^2}$ is the boost factor of the wall, and $+~(-)$ refers to fermions (bosons).

In \autoref{eq:force}, the CP even term is first order in derivatives, while the CP odd term is second order in derivatives; thus we can solve the CP even and odd parts separately. Following Ref.~\cite{Basler:2021kgq}, we introduce the following definition
%----
\begin{align}
\mu_i\equiv \mu_{i,1e}+\mu_{i,2o}+\mu_{i,2e},  \quad \quad \quad \delta f_i\equiv \delta f_{i,1e}+\delta f_{i,2o}+\delta f_{i,2e}.
\end{align}
%----
The evolution of $f_i$ is described by the Boltzmann equation
%----
\begin{align}
{\bf{L}}[f_i]\equiv \left ( v_g\partial_z+\dot{p}_z\partial_{p_z}\right )f_i=C[f_i],
\end{align}
%----
where ${\bf L}[f_i]$ is the Liouville operator and $v_g$ is the group velocity determined by WKB dispersion relation~\cite{Fromme:2006cm}
%----
\begin{align}
    v_g=\frac {P_z}{E_0}\left ( 1\pm\frac {\theta^\prime m^2}{2E^2_0E_{0z}} \right ).
\end{align}
%----
The $C[f_i]$ is a model-dependent collision integral associated with the interaction rate of the thermal bath~\cite{Cline:2000nw}. The terms in the fluid equation can be written as the average over-phase space of the form~\cite{Fromme:2006wx, Basler:2021kgq}
%----
\begin{align}
\left \langle X \right \rangle=\frac {\int {d^3p X(p)}}{\int {d^3p f^\prime_{0+}(m=0)}}, \quad\quad\quad \left \langle \frac {p_z}{E_0}X \right \rangle=\frac {\int {d^3p \frac {p_z}{E_0}X(p)}}{\int {d^3p f^\prime_{0+}(m=0)}},
\end{align}
%----
where $f^\prime_{0+}(m=0)$ can be written as
%----
\begin{align}
f^\prime_{0+}(m=0)\equiv f_i|_{\mathrm{fermion},\mu_i=0,\delta f_i=0,v_W=0}.
\end{align}
%----
Plasma velocities can be defined as
%----
\begin{align}
u_i\equiv \left \langle \frac{p_z}{E_0}\delta f_i \right \rangle.
\end{align}
%----

The second-order CP odd chemical potential is defined by the difference between the second-order chemical potential of the particle and its anti-particle, and a similar definition follows for corresponding plasma velocities,
%----
\begin{align}
\mu_{i,2}\equiv \mu_{i,2o}-{\bar{\mu}}_{i,2o}, \quad \quad \quad u_{i,2}\equiv u_{i,2o}-{\bar{u}}_{i,2o}.
\end{align}
%----
The zeroth and first momenta of the collision integral can be written in terms of inelastic rate $\Gamma_{\mathrm{inel}}$ and total interaction rate $\Gamma_{\mathrm{tot}}$ by \cite{Cline:2000nw}
%----
\begin{align}
\left \langle C[f_i] \right \rangle=\Gamma_{\mathrm{inel}}\sum \mu_i, \quad \quad \quad
\left \langle \frac {p_z}{E_0}C[f_i] \right \rangle=-\Gamma_{\mathrm{tot}}u.
\end{align}
%----

For the generation of the baryon asymmetry, the first step is to produce asymmetry in left-handed quarks. We consider the effects of the strong sphaleron process, $W$-scattering, top Yukawa interaction, helicity flip, and Higgs number violation with the rate of $\Gamma_{ss}$, $\Gamma_{W}$, $\Gamma_{y}$, $\Gamma_{m}$, and $\Gamma_{h}$ respectively. The last two processes are relevant only in the broken phase. The transport equation for chemical potentials of the left-handed top quark, the conjugate of the right-handed bottom quark, left-handed bottom quark, Higgs bosons, and the corresponding plasma velocities are given as follows ~\cite{Fromme:2006wx, Fromme:2006cm, Basler:2021kgq}:
\begin{itemize}
    \item Left-handed top quarks $(t)$
    \begin{align}
		0 =   & 3 \vw K_{1,t} \left( \partial_z \mu_{t,2} \right) + 3\vw K_{2,t} \left( \partial_z m_t^2 \right) \mu_{t,2} + 3 \left( \partial_z u_{t,2} \right) \notag
		\\ &- 3\Gamma_y \left(\mu_{t,2} + \mu_{t^c,2} + \mu_{h,2} \right) - 6\Gamma_M \left( \mu_{t,2} + \mu_{t^c,2} \right) - 3\Gamma_W \left( \mu_{t,2} - \mu_{b,2} \right) \notag
		\\ &- 3\Gamma_{ss} \left[ \left(1+9 K_{1,t} \right) \mu_{t,2} + \left(1+9 K_{1,b} \right) \mu_{b,2} + \left(1-9 K_{1,t} \right) \mu_{t^c,2} \right] \label{Eq:TransportEquations:mut} \,,\\
		S_t = & -3K_{4,t} \left( \partial_z \mu_{t,2}\right) + 3\vw \tilde{K}_{5,t} \left( \partial_z u_{t,2}\right) + 3\vw \tilde{K}_{6,t} \left( \partial_z m_t^2 \right) u_{t,2} + 3\Gamma_t^\mathrm{tot} u_{t,2} \label{Eq:TransportEquations:ut} \,.  \end{align}
	\item Charge conjugation of right-handed top quarks $(t^c)$
		\begin{align}
    	0=    & 3 \vw K_{1,t} \left( \partial_z \mu_{t^c,2} \right)  + 3\vw K_{2,t} \left( \partial_z m_t^2 \right)  \mu_{t^c,2} + 3 \left( \partial_z u_{t^c,2} \right) \notag                                                                                 \\
		      & - 3\Gamma_y \left(\mu_{t,2} + \mu_{b,2} + 2\mu_{t^c,2} + 2\mu_{h,2} \right) - 6\Gamma_M \left( \mu_{t,2} + \mu_{t^c,2} \right) \notag                                                                                                           \\
		      & - 3\Gamma_{ss} \left[ \left( 1+9 K_{1,t}\right) \mu_{t,2} + \left(1+9K_{1,b}\right) \mu_{b,2} + \left(1-9K_{1,t}\right) \mu_{t^c,2} \right] \label{Eq:TransportEquations:mutc} \,                                                              \\
		S_t = & -3K_{4,t} \left( \partial_z \mu_{t^c,2}\right) + 3\vw \tilde{K}_{5,t} \left( \partial u_{t^c,2}\right) + 3\vw \tilde{K}_{6,t} \left( \partial_z m_t^2\right) u_{t^c,2} + 3\Gamma_t^\mathrm{tot} u_{t^c,2} \label{Eq:TransportEquations:utc} \,.
	\end{align}
	\item Left-handed bottom quarks $(b)$
	\begin{align}
		0 =   & 3\vw K_{1,b} \left(\partial_z \mu_{b,2}\right) + 3 \left(\partial_z u_{b,2} \right) - 3\Gamma_y \left( \mu_{b,2} + \mu_{t^c,2} + \mu_{h,2} \right) - 3\Gamma_W \left( \mu_{b,2} - \mu_{t,2} \right) \notag \label{Eq:TransportEquations:mub}    \\
		      & - 3\Gamma_{ss} \left[ \left( 1 + 9K_{1,t}\right) \mu_{t,2} + (1+9K_{1,b}) \mu_{b,2} + (1-9K_{1,t}) \mu_{t^c,2} \right] \,                                                                                                                     \\
		0 =   & -3K_{4,b} \left( \partial_z \mu_{b,2} \right) + 3\vw \tilde{K}_{5,b} \left(\partial_z u_{b,2}\right) + 3\Gamma_b^\mathrm{tot} u_{b,2} \label{Eq:TransportEquations:ub} \,.
	\end{align}
	\item Higgs
	\begin{align}
		0 =   & 4\vw K_{1,h} \left( \partial_z \mu_{h,2}\right) +
		4\left( \partial_z u_{h,2}\right) - 3\Gamma_y \left(
		\mu_{t,2} + \mu_{b,2} + 2\mu_{t^c,2} + 2\mu_{h,2} \right) -
		4\Gamma_h
		\mu_{h,2} \label{Eq:TransportEquations:muh} \,,\\
		0 =   & -4K_{4,h} \left( \partial_z \mu_{h,2} \right) + 4\vw \tilde{K}_{5,h} \left( \partial_z u_{h,2} \right) + 4\Gamma_h^\mathrm{tot} u_{h,2} \label{Eq:TransportEquations:uh} \,,
	\end{align}

\end{itemize}
$S_t$ denotes the source term of the top quark that can be written as
%----
\begin{align}
S_t=-v_WK_{8,t}\partial_z(m^2_t \partial_z \theta)+v_WK_{9,t}(\partial_z \theta)m^2_t(\partial_z m^2_t).
\label{Eq:sourceterm}
\end{align}
%----
The source term for the bottom quark can be neglected due to the suppression factor $m_b^2/m_t^2 \sim 10^{-3}$.
Thermal transport coefficients are defined as
%----
\begin{subequations}
	\begin{align}
	    K_{1,i}&=-\left \langle \frac {p^2_z}{E_0^2}\partial_E^2f_{i,0} \right \rangle,  &K_{2,i}&=\left \langle \frac {\partial^2_Ef_{i,0}}{2E_0} \right \rangle,\\
K_{4,i}&=\left \langle \frac {p^2_z}{E_0^2}\partial_Ef_{i,0} \right \rangle,
&{\tilde{K}}_{5,i}&=\left [ \frac {p^2_z}{E_0^2}\partial_Ef_{i,0} \right ],\\
{\tilde{K}}_{6,i}&=\left [ \frac {E_0^2-p^2_z}{2E_0^3}\partial_Ef_{i,0} \right ],
&K_{8,i}&=\left \langle \frac {\left | p_z \right |\partial_Ef_{i,0}}{2E_0^2E_{0z}} \right \rangle,\\
K_{9,i}&=\left \langle \frac {|p_z|}{4E_0^3E_{0z}} \left ( \frac {\partial_E f_{i,0}}{E_0}-\partial^2_E f_{i,0} \right ) \right \rangle , \hspace{.8cm} &&
	\end{align}
	\label{Eq:Kfactors}
\end{subequations}
%----
with the expectation values given by
%----
\begin{align}
    \left \langle X \right \rangle=\frac {\int {d^3p X(p)}}{\int {d^3p \partial_Ef_{0+}(m=0)}}, \quad\quad
\left [ X \right ]=\frac {\int {d^3p X(p)}}{\int {d^3p f_{i,0,v_W}}}=\frac {\int {d^3p X(p)}}{\int {d^3p f_{i,0}|_{v_W=0}}},\\
\end{align}
%----
and the distribution function  defined as
%----
\begin{align}
    f_{i,0}=f_i|_{\mu_i=0,\delta f_i=0,v_w=0},\quad
f_{0+}= f_i|_{\mathrm{fermion},\mu_i=0,\delta f_i=0,v_w=0}, \quad
f_{i,0,v_w}=f_{i,0}+v_Wp_z\partial_{E_0}f_{i,0}.
\label{eq:distribution function}
\end{align}
%----

The third equation in \autoref{eq:distribution function} is a Taylor expansion; hence, it is valid only for small bubble wall velocity $v_W$. {We assume  $v_W=0.1$.} The transport equation with full dependence on the wall velocity is provided in Ref.~\cite{Cline:2020jre}. The values for the strong sphaleron rate, top Yukawa rate, Higgs number violating rate and rate for spin-helicity flipping rate for the top quark are given by \cite{Fromme:2006wx, Basler:2021kgq, Huet:1995sh, Moore:1997im}
%----
\begin{align}
    \Gamma_{ss}&=4.9\times 10^{-4}T\,,                &\Gamma_y&=4.2\times10^{-3}T\,,
    \nonumber \\
    \Gamma_m &=\frac {m^2_t(z,T)}{63T}\,,& \Gamma_h&=\frac {m^2_W(z,T)}{50T}\,,
     \label{eq:rates}
\end{align}
%----
where $z$ is the distance. The $W$ exchange rate can be approximated as the total Higgs interaction $\Gamma_W=\Gamma_h^{tot}$.

Finally, the asymmetry in left-handed quarks is converted into baryon asymmetry by electroweak sphaleron transition which can be calculated as~\cite{Cline:2000nw}
%----
\begin{align}
\eta_{B}=\frac {n_B}{s}=\frac {405 \Gamma_{ws}}{4\pi^2v_W g_\star T}\int_{0}^{\infty}dz\mu_{B_L}\exp\left ( -\frac {45\Gamma_{ws}z}{4v_W} \right ),
\label{eq:baryonasymmetry}
\end{align}
%----
where $\Gamma_{ws}\simeq 1\times 10^{-6}T$  is the weak sphaleron rate estimated by lattice calculation~\cite{Moore:2000mx} and $g_\star\simeq 106.75$ is the effective degrees of freedom at the electroweak scale.  The chemical potential for left-handed quarks $\mu_{B_L}$ is given by
%----
\begin{align}
\mu_{B_L}=\frac 12\left ( 1+4K_{1,t} \right )\mu_t+\frac 12\left ( 1+4K_{1,b} \right )\mu_b-2K_{1,t}\mu_{t^c}.
\label{eq:muBL}
\end{align}
%----
We solved the top transport equation and estimate the baryon asymmetry of the Universe with  {\bf{BSMPT v2}}~\cite{Basler:2020nrq}.\footnote{ We used CosmoTransitions to implement effective potential and calculate $T_c$, $T_n$, and tunneling profile as mentioned in \autoref{sec:EWPT_GW}. Currently, BSMPT does not provide a framework to compute $T_n$ and tunneling profile. We used $T_n$ and tunneling profile as input to BSMPT to solve the transport equation and compute the baryon asymmetry.}

%\footnote{\color{blue} The effective potential implementation in the BSMPT package and our implementation within CosmoTransition are almost identical except for a slight difference in the thermal function $J_{b/f}$. In this study, we use the implementation of CosmoTransitions as it accounts for higher-order terms.}

%%%%%%%%%%%%%%%%%%%%%%%%%%%%%%%%%%%%%%%%%%%%%%%%%%%%%%%%%%%
\section{Baryon Asymmetry in the C2HDM}
\label{sec:BassymC2HDM}
%%%%%%%%%%%%%%%%%%%%%%%%%%%%%%%%%%%%%%%%%%%%%%%%%%%%%%%%%%%

%----
\begin{figure}[b!]
    \centering
    \includegraphics[width=0.7\hsize]{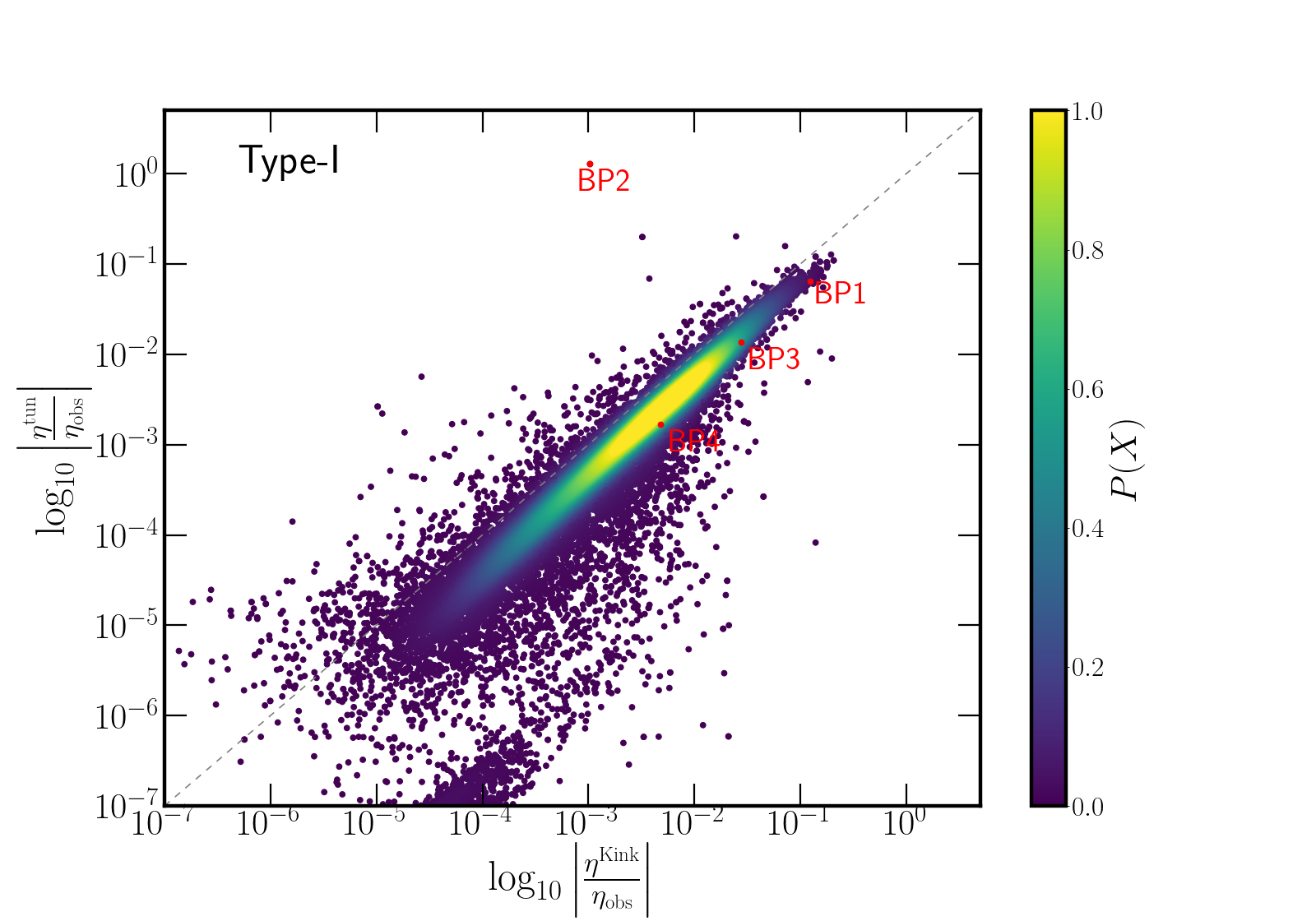}
    \caption{Comparison of the magnitude of baryon asymmetry computed using the profile from the solution of the tunneling equation and the kink profile at $T_n$ color coded with the probability distribution. The red color denotes the benchmark points presented in \autoref{tab:1}.}
    \label{fig:eta_comparison}
\end{figure}
%----

In this section, we estimate the baryon asymmetry generated via electroweak baryogenesis using the semi-classical force method. A key ingredient in this calculation is the estimation of the bubble profile. As discussed in \autoref{subsec:tunprof}, it is usual in the literature to parametrize bubble profile by the kink profile~\cite{Bodeker:2004ws, Fromme:2006wx, Fromme:2006cm, Basler:2021kgq}. In this section, in addition to deriving the BAU in the C2HDM framework, we pay close attention to the viability of the kink profile by comparing it with the bubble profile obtained by explicitly solving the tunneling equation using  {\tt CosmoTransitions}~\cite{Wainwright:2011kj}.

%----
\begin{figure}[t!]
\includegraphics[width=0.49\textwidth]{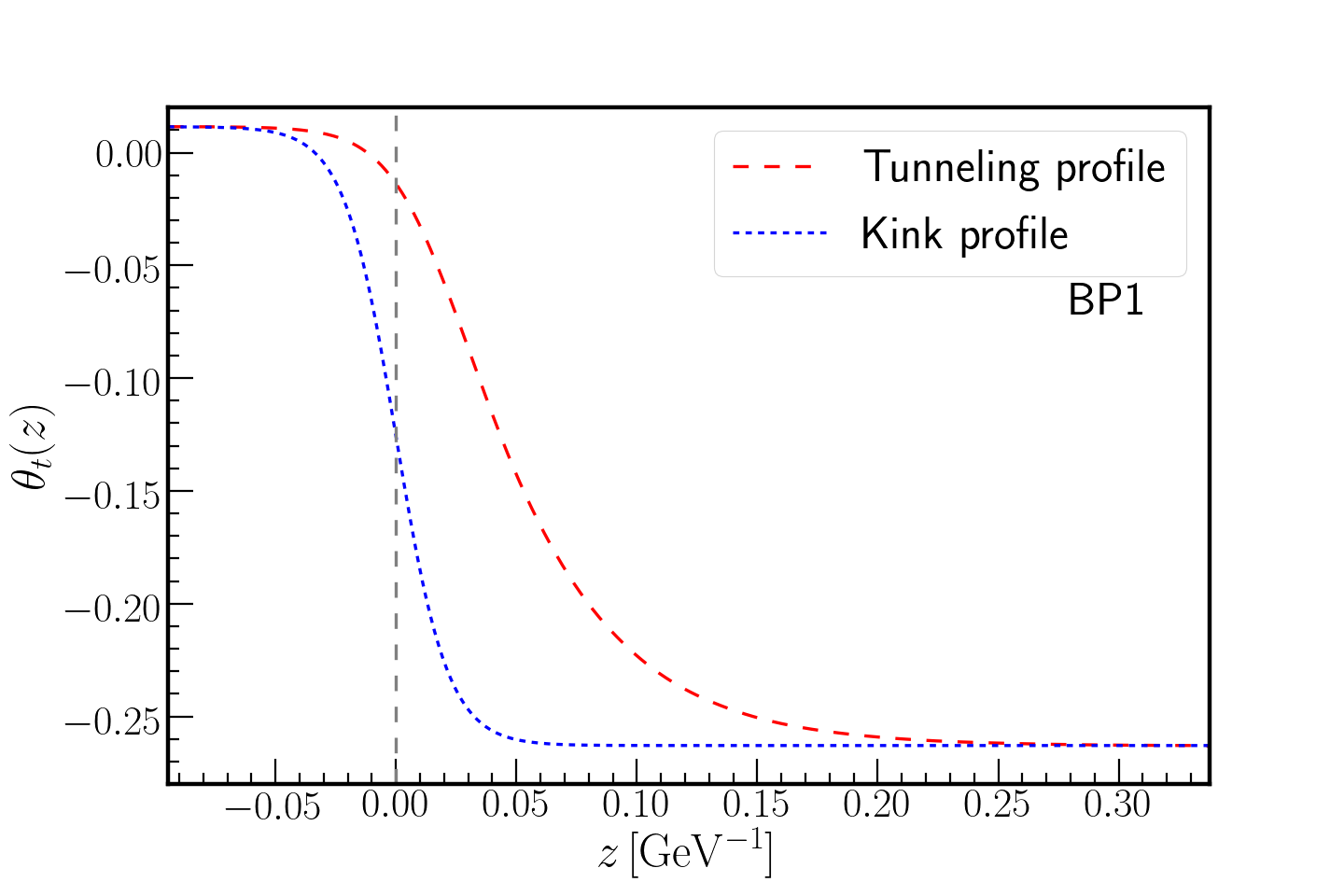}
\includegraphics[width=0.49\textwidth]{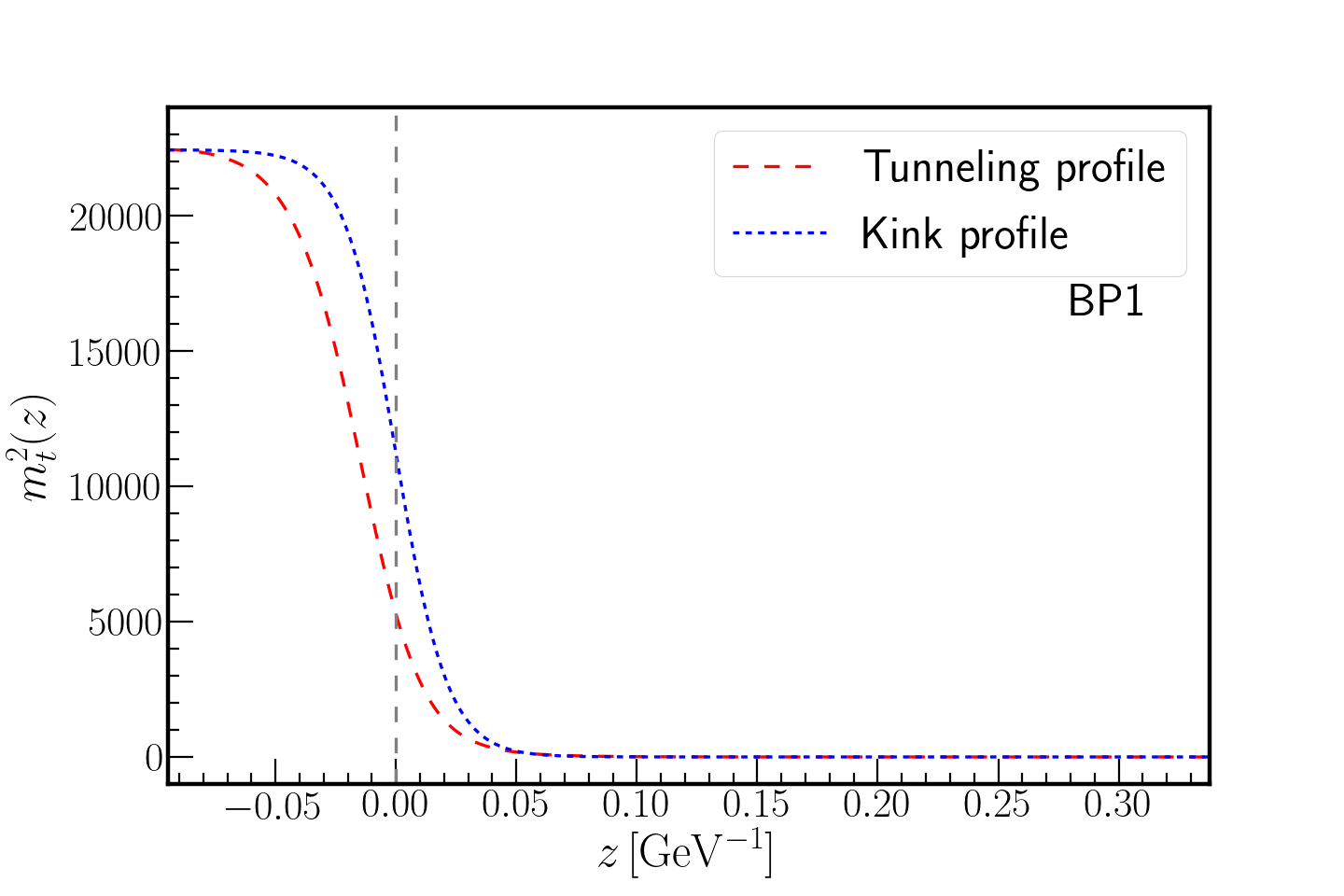}
\caption{Evolution of the phase of the top mass $\theta_t$ and the square of the top mass with the wall distance for the BP1, as defined in \autoref{tab:1}. The value of the top mass near the bubble wall is larger for the kink profile, thereby leading to a larger value of asymmetry. }
\label{BP1}
\end{figure}
%----

In \autoref{fig:eta_comparison}, we compare the magnitude of baryon asymmetry estimated using these two bubble profiles at the nucleation temperature with the color code representing the distribution probability. First, we observe that the C2HDM can satisfy the observed baryon asymmetry, matching the observed value $\eta_{obs}$. However, these points are rare in our parameter space scan. We highlight one of these points in \autoref{fig:eta_comparison} as benchmark point 2 (BP2)  and detailed define it in \autoref{tab:1}.\footnote{The VEV-insertion approximation (VIA)~\cite{Riotto:1995hh} has been found to produce baryon asymmetry values that are two to three orders of magnitude larger than the semi-classical force method adopted in the current study~\cite{Basler:2021kgq}, displaying a larger number of points satisfying $\eta_{obs}$. Several works have raised criticisms about the validity of the approximations used in this alternative method. One particular argument is that the expansion utilized in deriving the source term for the top quark in the VIA approach may encounter limitations due to the substantial mass of the top quark~\cite{Cline:2020jre,Basler:2021kgq}. It is important to highlight that recent improvements have been made in treating the source term in the VIA method~\cite{Postma:2022dbr}.} Second, we observe in \autoref{fig:eta_comparison} that for most of the points in the parameter space,  the kink profile solution leads to larger values than the profile obtained by solving the tunneling equation. In addition, there is a non-negligible fraction of points where the kink profile overestimates the asymmetries by a few orders of magnitude in comparison to the profile from the solution of the tunneling equation.

In most cases, we can understand the difference in asymmetry using two profiles by looking at the behavior of the source term~\autoref{Eq:sourceterm} in front of the bubble wall. Specifically, the sign of $\partial_z \theta_t$ in front of the bubble wall determines the sign of the source term $S_t$, thereby influencing the overall asymmetry. In most cases, a negative (positive) $\partial_z \theta_t$ results in a positive (negative) source term $S_t$, leading to a positive (negative) asymmetry. The kink profile typically provides a higher value for the top mass around the bubble wall, and thereby a higher magnitude for the source term in~\autoref{Eq:sourceterm}. This feature is illustrated in~\autoref{BP1} using our benchmark point 1 (BP1)  as defined in \autoref{tab:1}. Even when the change in phase of the top mass $\theta_t$ has a larger magnitude for the tunneling profile, the value of the top mass is higher for the kink profile, and subsequently, the kink profile has a larger asymmetry. Therefore, the behavior of the top mass is the dominant factor in estimating the magnitude of the asymmetry compared to the phase of top mass $\theta_t$.

%----
\begin{figure}[t!]
\includegraphics[width=0.49\textwidth]{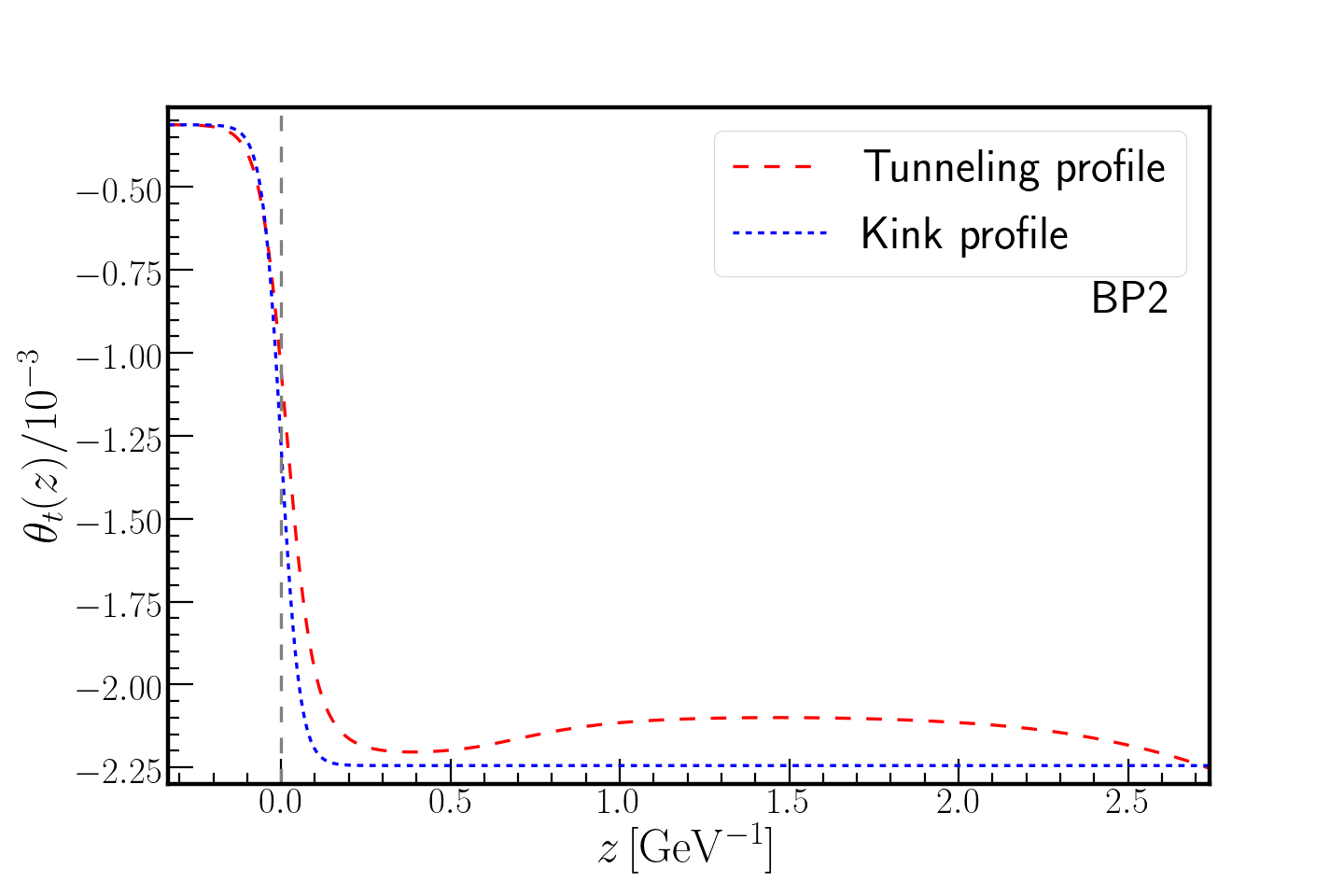}
\includegraphics[width=0.49\textwidth]{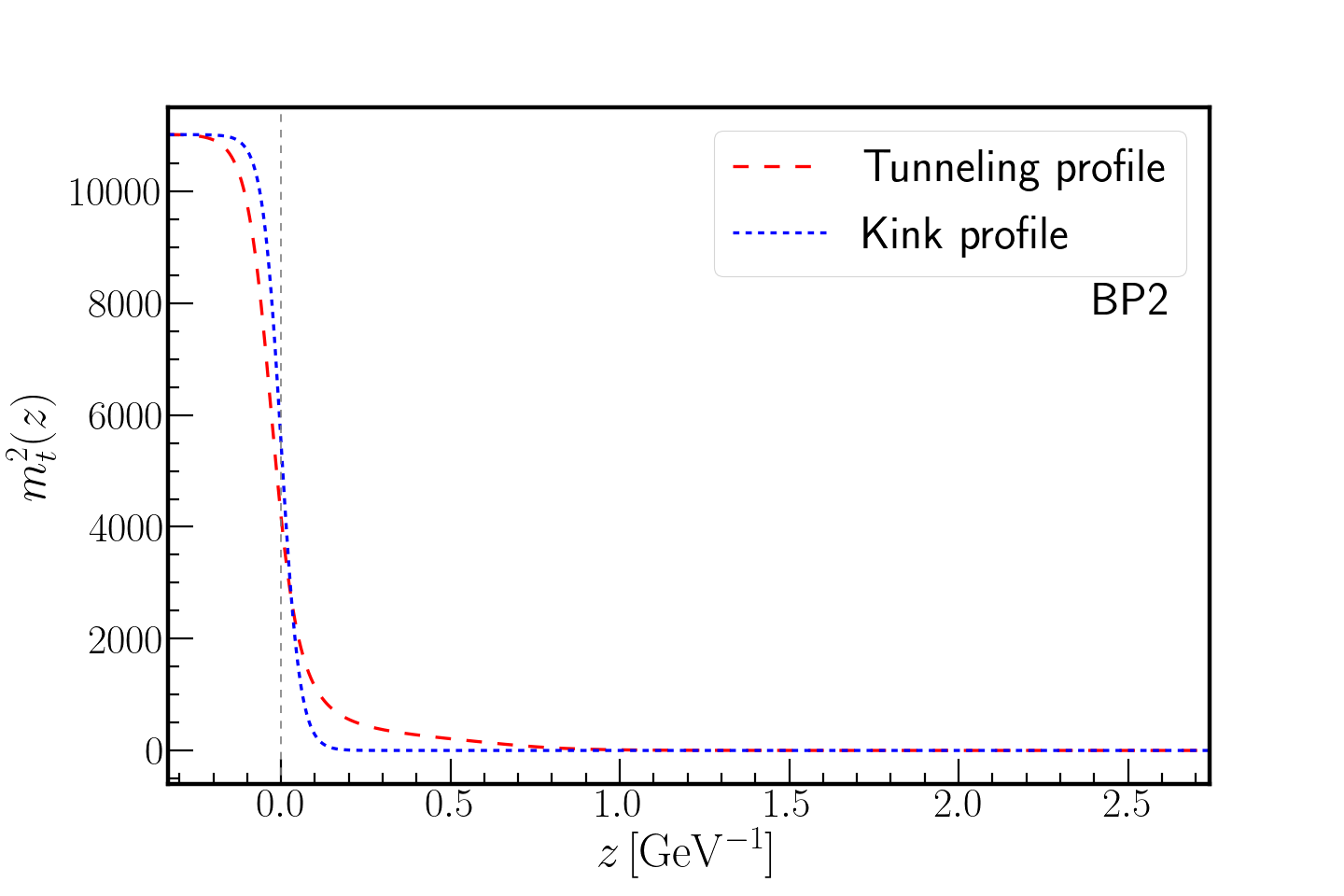}
\caption{The evolution of phase of top mass $\theta_t$ and square of top mass with the wall distance for the BP2. It takes a larger wall distance $z$ for the top mass value to drop to zero in the tunneling profile compared to the kink profile. This translates into the source term being active for a larger distance and, subsequently, the larger value of asymmetry for the tunneling profile compared to the kink profile. }
\label{BP2}
\end{figure}
%----

%----
\begin{figure}[t!]
\includegraphics[width=0.49\textwidth]{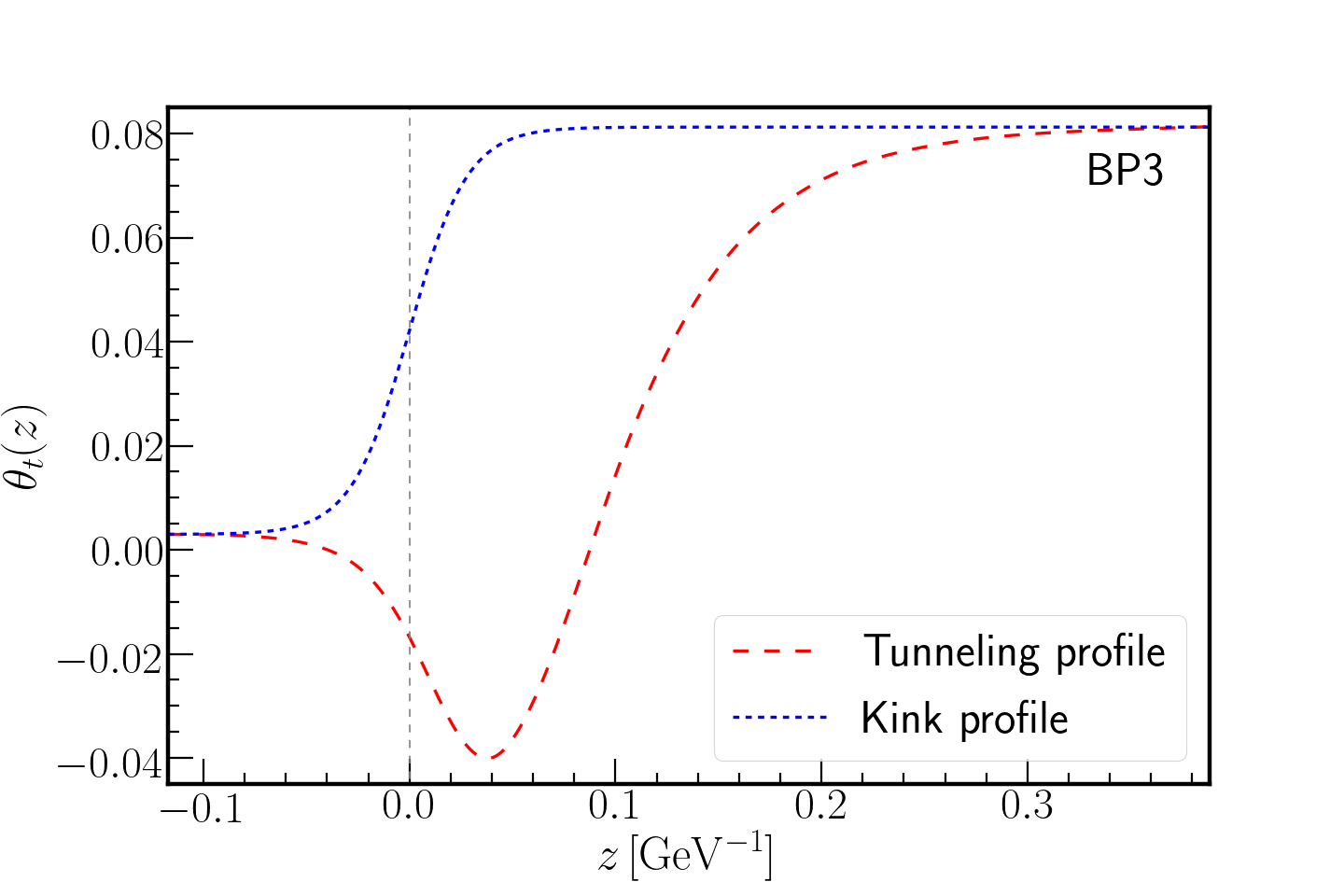}
\includegraphics[width=0.49\textwidth]{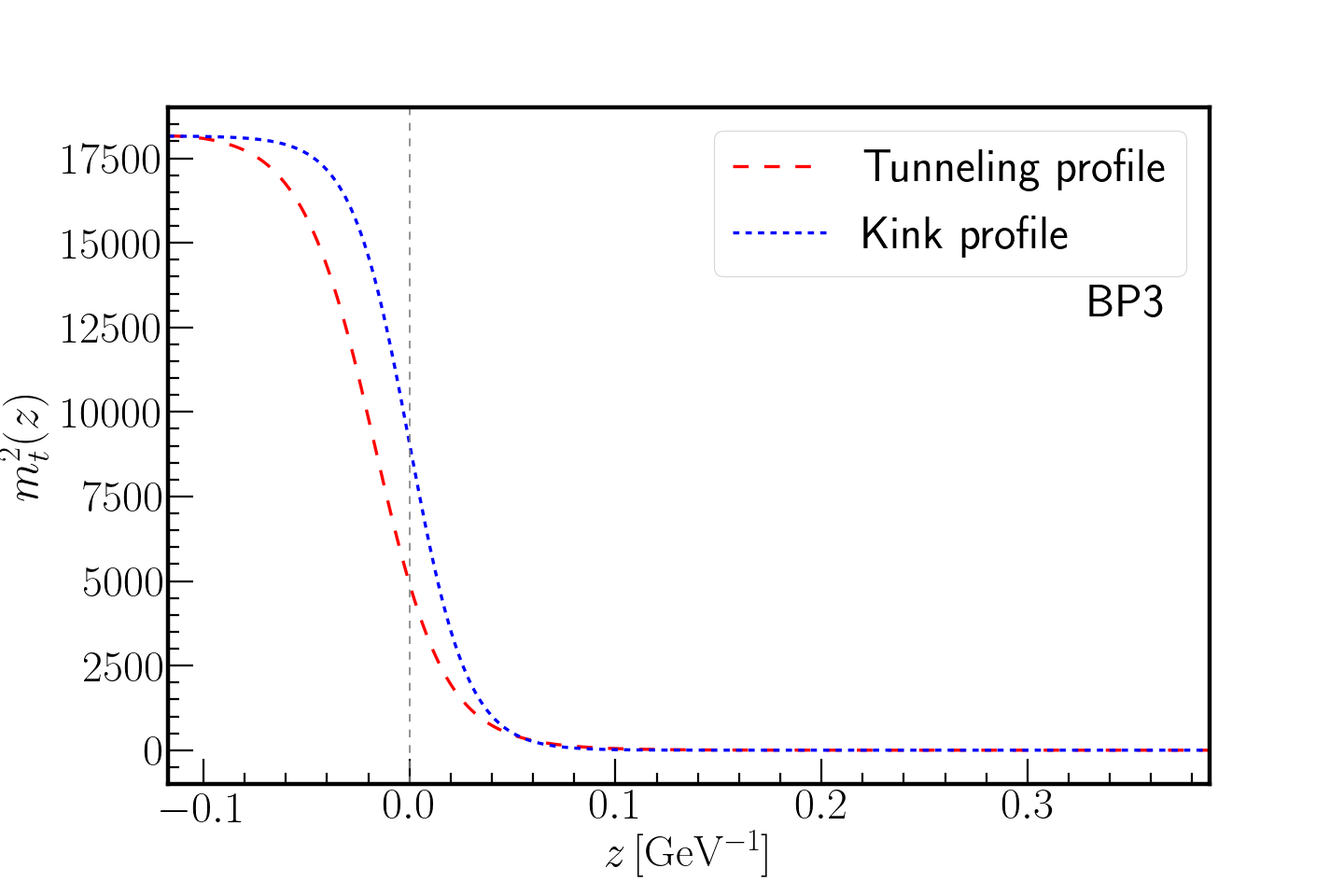}
\caption{The evolution of phase of top mass $\theta_t$ and square of top mass with the wall distance for the BP3. The phase of top mass near the bubble wall is an increasing function for the kink profile and decreasing function for the tunneling profile; subsequently, both asymmetries have opposite signs.}
\label{BP3}
\end{figure}
%----

There are instances where the top mass in the tunneling profile is smaller, but does not quickly drop to zero compared to the kink profile. This translates into the source term being active for a larger wall distance for the tunneling profile compared to the kink profile. In this case, the magnitude of asymmetry calculated using the tunneling profile would have a larger value compared to the kink profile. The above feature is illustrated in the case of BP2 shown in~\autoref{BP2}, where the asymmetry differs by two orders. Once again, we highlight that BP2 can explain the value of the observed baryon asymmetry $\eta_{obs}$ when using the explicit solution for the tunneling equation. The above characteristic of the tunneling profile will permit significant baryon asymmetry even though the change in CP violating phase is relatively small. \\

Finally, there are instances where even the sign of the derivative for the CP phase   $\partial_z\theta_t$ near the bubble wall differs between the kink and tunneling profiles. We illustrate this scenario with the benchmark point 3 (BP3) presented in~\autoref{BP3}. Near the bubble wall, $\theta_t$ exhibits an increasing trend for the kink profile, while it displays a decreasing trend for the explicit solution for the tunneling profile. Consequently, the asymmetry is positive for the tunneling profile and negative for the kink profile, despite both profiles having identical endpoints.  These findings emphasize the importance of accurately determining the bubble profile and highlight the discrepancies that can arise when relying on the kink profile approximation.

%%%%%%%%%%%%%%%%%%%%%%%%%%%%%%%%%%%%%%%%%%%%%%%%%%%%%%%%%%%
\section{Summary}
\label{sec:summary}
%%%%%%%%%%%%%%%%%%%%%%%%%%%%%%%%%%%%%%%%%%%%%%%%%%%%%%%%%%%

In this work, we explored the phase transition pattern and the feasibility of generating the observed baryon asymmetry of the Universe within the C2HDM framework while considering  the theoretical and experimental constraints. First, we carefully examined the essential elements in the shape of the Higgs potential, specifically focusing on the formation of the barrier and the upliftment of the true vacuum state. These factors are critical in facilitating the phase transition from a smooth crossover to a strong first order phase transition. We observe that the intensity of the phase transition is linked to the elevation of the true vacuum relative to the symmetric vacuum state at zero temperature~\cite{Dorsch:2017nza,EWPT-NMSSM,EWPT-Nature,Goncalves:2021egx}. This phenomenon occurs due to the prevalence of one-loop effects over thermal corrections, particularly when $\xi_c>1$~\cite{Goncalves:2021egx}. However, if the vacuum upliftment measure is too large, the universe becomes trapped in the false vacuum state, rendering no solution for the nucleation temperature~\autoref{eq:Tncond}. This leads to parameter points with $\Delta \mathcal{F}_0/ | \mathcal{F}_0^{\rm SM}| \gtrsim 0.87$ unphysical, which excludes most of the $\xi_c>2$ points~\cite{Goncalves:2021egx,Biekotter:2022kgf}. Therefore, in electroweak baryogenesis studies, it is crucial to look at the nucleation temperature $T_n$ and not just at the critical temperature $T_c$.

When it comes to gravitational wave signals, only a small fraction of the parameter points in the Strong First-Order Electroweak Phase Transition parameter space of the C2HDM  can be probed by LISA. However, among the accessible points, those with a higher value of the $\xi_c$ parameter display particularly strong gravitational wave signals. Notably, the Type I parameter space points generally offer more promising gravitational wave signals compared to the Type II parameter points in the C2HDM. These differences can be traced to the more stringent flavor constraints imposed on the Type-II scenario that shape the parameter space.

We note that the C2HDM can describe the observed baryon asymmetry $\eta_{obs}$, albeit for a limited set of parameter space points. One specific point, BP2, was highlighted as a benchmark that satisfied the observed asymmetry value. Furthermore, we contrast the impact on the baryon asymmetry calculation using two different approaches to describe the bubble profile,  namely  the usually adopted kink profile parameterization and the explicit solution for the tunneling equation. Our objective was to access the dependency of the resulting value of $\eta_B$ on these two approaches and evaluate their respective contributions to the baryon asymmetry calculation. We found that the majority of points in our parameter space scan yield similar results from both approaches. Nonetheless, a non-negligible portion of points exhibits significant discrepancies between these two methods. Specifically, the kink profile approximation often displays higher asymmetry values compared to the explicit solution obtained from the tunneling equation. In some cases, the discrepancy was by several orders of magnitude. The difference in the asymmetry value for the two profiles was scrutinized in terms of the behavior of the source term in front of the bubble wall.

Undoubtedly, the task of achieving a baryon asymmetry of the universe  that aligns with the observed value poses significant challenges. The requirements of a strong first-order electroweak phase transition, substantial CP violation, and stringent theoretical and experimental constraints make the generation of a compatible BAU a formidable task. However, the discrepancies observed in calculations performed using different profile assumptions provide avenues for improving the accuracy of computing the BAU. The comparison between the kink profile and the explicit solution for the tunneling profiles provided valuable insights into the estimation of baryon asymmetry, emphasizing the importance of accurately determining the bubble profile for a more robust analysis of electroweak baryogenesis.

%%%%%%%%%%%%%%%%%%%%%%%%%%%%%%%%%%%%%%%%%%%%%%%%%%%%%%%%%%%
\section*{Acknowledgements}
\label{sec:acknowledgements}
%%%%%%%%%%%%%%%%%%%%%%%%%%%%%%%%%%%%%%%%%%%%%%%%%%%%%%%%%%%
We would like to thank Margarete M\"{u}hlleitner and Jonas Wittbrodt for useful discussions about BSMPT and ScannerS, as well as Peter Athron for useful discussion about percolation. DG, AK, and YW thank the U.S.~Department of Energy for the financial support, under grant number DE-SC 0016013. Some computing for this project was performed at the High Performance Computing Center at Oklahoma State University, supported in part through the National Science Foundation grant OAC-1531128.

%%%%%%%%%%%%%%%%%%%%%%%%%%%%%%%%%%%%%%%%%%%%%%%%%%%%%%%%%%%
\appendix
\section{Parametrization for C2HDM Scan}
\label{app:parameters}
{In this appendix, we present the detailed parameterization for the C2HDM adopted in our parameter space scan in~\autoref{eq:param_scan}. From the following minimization conditions at zero temperature}
\begin{subequations}
\label{equ:Umin}
\begin{align}
    \frac{dV}{d\omega_1} &= \frac{\omega_1\left(2m_{11}^2+\lambda_1\omega_1^2+(\lambda_5^rc_{2\omega_\theta}-\lambda_5^is_{2\omega_\theta})\omega_2^2 + (\lambda_3+\lambda_4)\omega_2^2\right)-2m_{12}^2\omega_2c_{\omega_\theta}}{2}=0,\\
    \frac{dV}{d\omega_2} &= \frac{\omega_2\left(2m_{22}^2+\lambda_2\omega_2^2+(\lambda_5^rc_{2\omega_\theta}-\lambda_5^is_{2\omega_\theta})\omega_1^2 + (\lambda_3+\lambda_4)\omega_1^2\right)-2m_{12}^2\omega_1c_{\omega_\theta}}{2}=0,\\
    \frac{dV}{d\omega_\theta} &= \frac{\omega_1\omega_2\left(2m_{12}^2s_{\omega_\theta} - \omega_1\omega_2(\lambda_5^rs_{2\omega_\theta} + \lambda_5^ic_{2\omega_\theta})\right) }{2} = 0,
\end{align}
\end{subequations}
we can write the  tree-level parameters as
\begin{subequations}
\label{equ:VacCond_app}
\begin{align}
    m_{11}^2 &= \frac{m_{12}^2t_\beta c_\theta}{c_{2\theta}} - \frac{1}{2}v^2\left(c_\beta^2\lambda_1 + s_\beta^2\left(\lambda_3+\lambda_4+\lambda_5^r/c_{2\theta}\right)\right), \\
    m_{22}^2 &= \frac{m_{12}^2c_\theta}{t_\beta c_{2\theta}} - \frac{1}{2}v^2\left(s_\beta^2\lambda_2 + c_\beta^2\left(\lambda_3+\lambda_4+\lambda_5^r/c_{2\theta}\right)\right), \\
    \lambda_5^i &= \frac{2m_{12}^2s_\theta}{s_\beta c_\beta c_{2\theta}v^2} - \lambda_5^r t_{2\theta}.
\end{align}
\end{subequations}
Note that in the limit $\theta\to 0$, where $\alpha_c$ also goes to zero, we recover the CP-conserving 2HDM.

{From the quadratic terms in the potential, we have the following relations for the charged scalar mass and neutral scalar mass matrix:}
\begin{subequations}
\begin{align}
    m_{H^\pm}^2 &= \frac{m_{12}^2c_\theta}{s_\beta c_\beta c_{2\theta}} - \frac{1}{2}v^2 \left(\lambda_4 + \lambda_5^r/c_{2\theta}\right),\\
    \mathcal{M}^2_{N} &= \left(\begin{array}{ccc}
        \mathcal{M}_{11}^2 & \mathcal{M}_{12}^2 & \mathcal{M}_{13}^2 \\
        \mathcal{M}_{21}^2 & \mathcal{M}_{22}^2 & \mathcal{M}_{23}^2 \\
        \mathcal{M}_{31}^2 & \mathcal{M}_{32}^2 & \mathcal{M}_{33}^2
    \end{array}\right),\\
    \mathcal{M}_{11}^2 & = m_{12}^2t_\beta c_\theta + \lambda_1v^2c_\beta^2, \\
    \mathcal{M}_{22}^2 &= \frac{m_{12}^2c_\theta}{t_\beta} + \lambda_2v^2s_\beta^2,\\
    \mathcal{M}_{33}^2 &= \frac{m_{12}^2}{2s_\beta c_\beta c_{2\theta}}(3c_\theta - c_{3\theta}) - \frac{\lambda_5^rv^2}{c_{2\theta}},\\
    \mathcal{M}_{12}^2 &= \mathcal{M}_{21}^2 = \frac{1}{2}\left(\frac{m_{12}^2(c_{3\theta}-3c_\theta)}{c_{2\theta}} + s_{2\beta}\left(\lambda_3+\lambda_4 + \lambda_5^r/c_{2\theta}\right)v^2\right), \\
    \mathcal{M}_{13}^2 &= \mathcal{M}_{31}^2 = -\frac{m_{12}^2s_\theta}{c_\beta}, \\
    \mathcal{M}_{23}^2 &= \mathcal{M}_{32}^2 = -\frac{m_{12}^2s_\theta}{s_\beta}.
\end{align}
\end{subequations}
From $m_{H^\pm}^2$ and $\mathcal{M}$, we  obtain the expressions for $\lambda_{1,\cdots,4},\lambda_5^r$ in terms of the physical parameters in~\autoref{equ:inputs}:
\begin{subequations}
\label{equ:lambdas_app}
\begin{align}
    \lambda_1 &= \frac{\sum_i O_{i1}^2m_i^2}{c_\beta^2 v^2} - \frac{m_{12}^2t_\beta c_\theta}{c_\beta^2 v^2}, \\
    \lambda_2 &= \frac{\sum_i O_{i2}^2m_i^2}{s_\beta^2 v^2} - \frac{m_{12}^2c_\theta}{t_\beta s_\beta^2 v^2}, \\
    \lambda_3 &= \frac{\sum_i O_{i1}O_{i2}m_i^2}{s_\beta c_\beta v^2} - \frac{m_{12}^2c_\theta}{s_\beta c_\beta v^2} + \frac{2m_{H^\pm}^2}{v^2}, \\
    \lambda_4 &= \frac{\sum_i O_{i3}^2m_i^2}{v^2} + \frac{m_{12}^2c_\theta}{s_\beta c_\beta v^2} - \frac{2m_{H^\pm}^2}{v^2},\\
    \lambda_5^r &= -\frac{c_{2\theta}\sum_i O_{i3}^2m_i^2}{v^2} + \frac{m_{12}^2(3c_\theta-c_{3\theta})}{s_{2\beta}v^2},
\end{align}
\end{subequations}
where $O$ is the rotation matrix in~\autoref{eq:betamixing} that diagonalizes $\mathcal{M}_N^2$, and $m_i^2$ for $i=1,2,3$ are the mass eigenvalues of $\mathcal{M}_N^2$ and can be identified with $m_h$, $m_{H_\uparrow}$ and $m_{H_\downarrow}$.
{Note that the determinant of matrix $O$ in~\autoref{eq:betamixing} is $-1$, which is chosen such that the definition of $\alpha$ follows the same convention as the counterpart in CP-conserving 2HDM.}\footnote{To match the convention in {\tt ScannerS}, extra permutations and multiplications will be added to the rotation matrix.}
The parameters $\alpha_b$ and $\theta$ can be obtained by using $\mathcal{M}_{13}^2$ and $\mathcal{M}_{23}^2$,
\begin{subequations}
\label{equ:CPVangles_app}
\begin{align}
    s_{\alpha_b} &= \frac{s_{2\alpha_c}(m_2^2-m_3^2)}{2(m_1^2-m_2^2s_{\alpha_c}^2-m_3^2c_{\alpha_c}^2)t_{\alpha+\beta}}, \\
    s_\theta &= -\frac{c_\beta}{m_{12}^2}\sum_i m_i^2 O_{i1}O_{i3}.
\end{align}
\end{subequations}
With the three mixing angles $\alpha$, $\alpha_b$, and $\alpha_c$, we can evaluate the rotation matrix $O$ in \autoref{eq:betamixing} and subsequently obtain $\lambda$'s using \autoref{equ:lambdas_app}. In this parametrization we choose $\beta$, $\alpha$ and $\alpha_c$ as independent parameters. The remaining $m_{11}^2$, $m_{22}^2$ and $\lambda_5^i$ can be calculated using \autoref{equ:VacCond_app}.
%%%%%%%%%%%%%%%%%%%%%%%%%%%%%%%%%%%%%%%%%%%%%%%%%%%%%%%%%%%
\section{ Gravitational Wave Signatures}
\label{app:GW}
First-order phase transition (FOPT) in the early universe can give rise to detectable stochastic GW signals today~\cite{Caprini:2015zlo, Cai:2017cbj,Caprini:2018mtu}. During FOPT,  GWs are generated by bubble collision, sound waves, and magnetohydrodynamic (MHD) turbulence. The GW spectrum $\Omega_{\mathrm{GW}}(f)$ is the current GW energy density per logarithmic frequency interval per critical energy density of the universe. That is $\int \Omega_{\mathrm{GW}}d\ln f$ is the fraction of GW energy density compared to the critical energy density of the universe. The GW energy density can be linearly approximated as
%-------
\begin{equation}   \Omega_{\mathrm{GW}}h^2=\Omega_{\mathrm{coll}}h^2+\Omega_{\mathrm{SW}}h^2+\Omega _{\mathrm{turb}}h^2\,.
\label{eq:GW}
\end{equation}
%-------
Here, $h=H_0/100~\text{Km}/ \text{s}/\text{Mpc}$ is the dimensionless Hubble parameter as it stands today. In the following, we discuss each of the components of \autoref{eq:GW}.

%-------
\subsection{Bubble collisions}

Nucleated bubbles undergo collisions that disrupt their spherical symmetry and give rise to gravitational waves. In the $\frac {\beta}{H_{n}}\gg 1$ limit, this can be described well by the thin wall and envelope approximation~\cite{Kosowsky:1991ua,Kosowsky:1992vn}. In this approximation, GW is generated mainly from envelopes, and the overlapped region is neglected. The gravitational wave spectrum from bubble collision can be modeled by~\cite{Jinno:2016vai}
%-------
\begin{equation}
     \Omega_{\mathrm{coll}}h^2=\frac {1.67 \times 10^{-5}}{(\beta/{H_{\star}})^2}\left ( \frac {k_c \alpha}{1+\alpha} \right )^{2}\left ( \frac {100}{g_{*}} \right )^{1/3}\left ( \frac {0.11v_w^3}{0.42+v_w^2} \right )\frac {3.8(f/f_{env})^{2.8}}{1+2.8(f/f_{env})^{3.8}}\,,
\end{equation}
%-------
where $k_c$ is the efficiency factor for the bubble collision~\cite{Kamionkowski:1993fg}
%-------
\begin{equation}
    k_c=\frac {0.715\alpha+\frac {4}{27}\sqrt{\frac{3\alpha}{2}}}{1+0.715\alpha}.
\end{equation}
%-------
Taking into account the redshift of the frequency, the peak frequency of the GW spectrum from bubble collision is given by~\cite{Huber:2008hg}
%-------
\begin{equation}
   f_{\mathrm{env}}=\left ( \frac {0.62 \times \beta/H_{\star}}{1.8-0.1v_w+v_w^2} \right )16.5\times 10^{-6}\left ( \frac {T_n}{100 ~\rm GeV} \right )\left ( \frac {g_{*}}{100} \right )^{1/6} \text{Hz},
\end{equation}
%-------
where first term in right side represents the peak frequency of the GW spectrum from bubble collision at $T_n$.

%-------
\subsection{Sound waves}
The energy released from the phase transition to the plasma can either go to heat or fluid motion. Numerical estimation shows that the energy-momentum tensor of the fluid after bubble collision is similar to that of an ensemble of sound waves~\cite{Hindmarsh:2017gnf}. Remarkably, these sound waves serve as a notable source of gravitational waves.  The gravitational wave spectrum stemming from these sound waves can be modeled by~\cite{Hindmarsh:2013xza, Hindmarsh:2017gnf, Hindmarsh:2016lnk}
%-------
\begin{equation}
     \Omega_{\mathrm{SW}}h^2=\frac {2.65 \times 10^{-6}}{\beta/H_{\star}}\left ( \frac {k_s \alpha}{1+\alpha} \right )^2\left ( \frac {100}{g_{*}} \right )^{1/3}v_w \left ( \frac {f}{f_{\mathrm{SW}}} \right )^3\left ( \frac {7}{4+3(f/f_{\mathrm{SW}})^2} \right )^{7/2}\Upsilon (\tau_{\mathrm{SW}}),
\end{equation}
%-------
where the peak frequency of the GW spectrum, accounting for the  redshift factor, can be written as~\cite{Huber:2008hg},
%-------
\begin{equation}
    f_{\mathrm{SW}}=\frac {1.9\times 10^{-5}}{v_w}\left ( \frac {\beta}{H_{\star}} \right )\left ( \frac {T_n}{100 ~\rm GeV} \right )\left ( \frac {g_{*}}{100} \right )^{1/6}\text {Hz}\,.
\end{equation}
%-------

For the radiation-dominant universe, the lifetime of the sound waves is finite. Recent studies ~\cite{Guo:2020grp, Hindmarsh:2020hop} show a suppression factor $\Upsilon (\tau_{\mathrm{SW}})$ due to the finite active period $\tau_{\mathrm{SW}}$ of sound waves
%-------
\begin{equation}
    \Upsilon (\tau_{\mathrm{SW}})=1-\frac {1}{\sqrt{1+2\tau_{\mathrm{SW}} H_{\star}}}\,,
\end{equation}
%-------
where the efficiency factor for the sound wave spectrum $k_s$ is given by ~\cite{Caprini:2015zlo}
%-------
\begin{equation}
    k_s=\frac {\alpha}{0.73+0.083\sqrt{\alpha}+\alpha}\,.
\end{equation}
%-------
The sound wave lasts until the turbulence starts to develop. The duration $\tau_{\mathrm{SW}}$ is quantified with~\cite{Hindmarsh:2017gnf}
%-------
\begin{equation}
    \tau_{\mathrm{SW}}=\frac {R_{\star}}{\overline{U_f}},
\end{equation}
%-------
where $R_{\star}\simeq (8\pi)^{1/3}v_w/\beta$ denotes the mean bubble separation and $\overline{U_f}^2$ represents the mean square velocity~\cite{Bodeker:2017cim}
%-------
\begin{equation}
    {\overline{U_f}}^2=\frac 34 \frac {\alpha}{1+\alpha}k_s.
\end{equation}
%-------

%--------
\subsection{MHD turbulence}

The energy injected into the plasma can induce turbulence in the fluid if the early universe plasma has an extremely high Reynolds number~\cite{Kamionkowski:1993fg}. This turbulent motion can be a source of gravitational waves~\cite{Witten:1984rs}. Additionally, a fully ionized plasma can give rise to a turbulent magnetic field under turbulent motion, leading to GWs. The GW spectrum stemming from turbulence can be parametrized as~\cite{Caprini:2009yp}
%-------
\begin{equation}
   \Omega _{\mathrm{turb}}h^2=\frac {3.35 \times 10^{-4}}{\beta/H_{\star}}\left ( \frac {k_t \alpha}{1+\alpha} \right )^{3/2}\left ( \frac {100}{g_{*}} \right )^{1/3}v_w \left ( \frac {f}{f_{\mathrm{SW}}} \right )^3\frac {1}{\left [1+(f/f _{\mathrm{turb}})  \right ]^{11/3}(1+8\pi f/h_{\star})},
\end{equation}
%-------
where $h_{\star}$ is the factor accounting redshift of the frequency
%-------
\begin{equation}
    h_{\star}=16.5\times 10^{-6}\left ( \frac {T_n}{100~ \rm GeV} \right )\left ( \frac {g_{*}}{100} \right )^{1/6} \text {Hz}.
\end{equation}
%-------
Based on insights drawn from numerical simulation, we adopt $k _{\mathrm{turb}}=0.05k_s$~\cite{Caprini:2015zlo}. Finally, the peak frequency for GW spectrum arising from MHD turbulence can be written as~\cite{Caprini:2009yp},
%-------
\begin{equation}
    f _{\mathrm{turb}}=\frac {2.7\times 10^{-5}}{v_w}\left ( \frac {\beta}{H_{\star}} \right )\left ( \frac {T_n}{100~\rm GeV} \right )\left ( \frac {g_{*}}{100} \right )^{1/6} \text{Hz}.
\end{equation}
%-------
%\end{enumerate}

%\end{itemize}
%%%%%%%%%%%%%%%%%%%%%%%%%%%%%%%%%%%%%%%%%%%%%%%%%%%%%%%%%%%
\section{Thermal Debye mass}
\label{app:Debye}

The thermal Debye masses $\overline{m}_{i}$ are given by~\cite{Arnold:1992rz,Basler:2016obg}
\begin{equation}
\overline{m}^2_{i}=m^2_i+\Pi_{ii}.
\end{equation}

Within  the context of the C2HDM, the self-energy corrections are written as follows~\cite{Comelli:1996vm}:
\begin{align}
\Pi_{\Phi_1\Phi_1}&=\left\{\begin{matrix}
\frac {1}{48}\left ( 12{\lambda_1}+8 \lambda_3+4 \lambda_4 +3(3g^2+g^{\prime 2}) \right )T^2, &\text{Type-I}, \\
\frac {1}{48}\left ( 12{\lambda_1}+8 \lambda_3+4 \lambda_4 +3(3g^2+g^{\prime 2})+12y_b^2 \right )T^2,  & \text{Type-II},
\end{matrix}\right.\\
\Pi_{\Phi_2\Phi_2}&=\left\{\begin{matrix}
\frac {1}{48}\left ( 12{\lambda_2}+8 \lambda_3+4 \lambda_4 +3(3g^2+g^{\prime 2})+12y_t^2+12y_b^2 \right )T^2, &\text{Type-I}, \\
\frac {1}{48}\left ( 12{\lambda_2}+8 \lambda_3+4 \lambda_4 +3(3g^2+g^{\prime 2})+12y_t^2 \right )T^2, & \text{Type-II},
\end{matrix}\right.\\
    \Pi_{W^a W^a}&=2g^2 T^2,\\
\Pi_{BB}&=2g^{\prime 2} T^2.
\end{align}

%%%%%%%%%%%%%%%%%%%%%%%%%%%%%%%%%%%%%%%%%%%%%%%%%%%%%%%%%%%
\bibliographystyle{refs}
\bibliography{references}

\end{document}